\documentclass[prd,onecolumn,showpacs,floatfix,amsmath,amssymb,floatfix]{revtex4}
\usepackage{graphicx,dcolumn,booktabs,bm}
\usepackage{longtable,lscape}
\usepackage{txfonts}
\usepackage{overpic}
\usepackage{multirow}
\usepackage{amssymb}
\usepackage{mathrsfs}
\usepackage{indentfirst}
\usepackage{cases}
\usepackage{color,ulem}
\usepackage{graphicx}
\usepackage{hyperref}

\begin{document}
%\begin{CJK}{GBK}{

\title{Axial charges of the hidden-charm pentaquark states}
\author{Guang-Juan Wang$^{1}$}\email{wgj@pku.edu.cn}
\author{Zhan-Wei Liu$^{2,3}$}\email{zhan-wei.liu@adelaide.edu.au}
\author{Shi-Lin Zhu$^{1,4}$}\email{zhusl@pku.edu.cn}

\affiliation{
$^1$School of Physics and State Key Laboratory of Nuclear Physics and Technology and Center of High Energy Physics, Peking University, Beijing 100871, China\\
$^2$School of Physical Science and Technology, Lanzhou University, Lanzhou
730000, China\\
$^3$Special Research Center for the Subatomic Structure of Matter (CSSM), Department of Physics, University of Adelaide, Adelaide SA 5005, Australia\\
$^4$Collaborative Innovation Center of Quantum Matter, Beijing
100871, China}
\date{\today}

\begin{abstract}

With the chiral quark model, we have calculated the axial charges of
the pentaquark states with $(I,I_3)=(\frac{1}{2},\frac{1}{2})$ and
$J^{P}=\frac{1}{2}^{\pm},\frac{3}{2}^{\pm}$, and $\frac{5}{2}^{\pm}$. The
$P_c$ states with the same $J^P$ quantum numbers but different
color-spin-flavor configurations have very different axial charges,
which encode important information on their underlying structures.
For some of the $J^{P}=\frac{3}{2}^{\pm}$ or $\frac{5}{2}^{\pm}$ pentaquark
states, their axial charges are much smaller than that of the
proton.

\end{abstract}


\maketitle

%%%%%%%%%%%%%%%%%%%%%%%%%%%%
\section{Introduction}\label{sec1}
%%%%%%%%%%%%%%%%%%%%%%%%%%%%

The LHCb Collaboration observed two hidden-charm pentaquark states,
$P_c(4380)$ and $P_c(4450)$, in the $\Lambda^0_b\to J/\psi p K^-$
process last year \cite{Aaij:2015tga}. The mass and width of the
lower state are $4380\pm8\pm29$ and $205\pm18\pm86$ MeV,
respectively, and its quantum number is probably $J^P=\frac{3}{ 2}^-$.
The mass and width of the higher state are $4449.8\pm1.7\pm2.5$ and $39\pm5\pm19$ MeV, respectively, and its quantum number is probably $J^P=\frac{5}{ 2}^+$.

Before the discovery of the two $P_c$ states, there existed
extensive theoretical investigations of the possible hidden-charm
pentaquark states
\cite{Yang:2011wz,Wu:2010jy,Uchino:2015uha,Karliner:2015ina}. Up to
now, the possible interpretations of these two hidden-charm
pentaquark states include the hidden-charm molecular pentaquarks
composed of an anticharmed meson and a charmed baryon
\cite{Yang:2011wz,Chen:2015loa, He:2015cea,
Burns:2015dwa,Wang:2015qlf,Lu:2016nnt,Shimizu:2016rrd,Shen:2016tzq,Wang:2016dzu,Huang:2015uda,Yamaguchi:2016ote,Ortega:2016syt,Yang:2015bmv},
the dynamically generated states
\cite{Wu:2010jy,Uchino:2015uha,Roca:2015dva,Xiao:2015fia,Chen:2015sxa,Wang:2015pcn},
the compact pentaquark states
\cite{Karliner:2015ina,Chen:2015moa,Maiani:2015vwa,Lebed:2015tna,Li:2015gta,Ghosh:2015ksa,Wang:2015epa,Wang:2015ava,Cheng:2015cca,Lebed:2015dca,Zhu:2015bba,Gerasyuta:2015djk,Wang:2016dzu,Chen:2016otp,Kopeliovich:2016gra,Ali:2016dkf},
and even a soliton \cite{Scoccola:2015nia}. Especially the higher
state was suggested as a $p$-wave excitation of the broad $s$-wave state
around 4450 MeV \cite{Wang:2015qlf,Chen:2016qju,He:2016pfa}. The
$P_c(4450)$ state was also speculated to arise from the nonresonant
signals from either the threshold effects or the triangle singularity
\cite{Guo:2015umn,Liu:2015fea,Mikhasenko:2015vca,Meissner:2015mza,Szczepaniak:2015hya,Carames:2016xcn,Guo:2016bkl}.
An extensive review of the hidden-charm pentaquark states can be
found in Ref. \cite{Chen:2016qju}. Recently, the
LHCb Collaboration confirmed these two $P_c$ states in the $\Lambda^0_b \to J/\psi p
\pi^-$ process in Ref. \cite{Aaij:2016ymb}, but once again the possibility that the narrow state would be
due to a triangle singularity was raised in Ref. \cite{Guo:2016bkl}. A revision of Refs. \cite{Guo:2015umn,Liu:2015fea,Guo:2016bkl} has been done in Ref. \cite{Bayar:2016ftu}, where it is shown that if the
quantum numbers of this state are those preferred by the experiment,
the triangle mechanism requires $\chi_{c1} p \rightarrow J/{\psi} p$ scattering with
$\chi_{c1} p$ in $P$ or $D$ waves at threshold and the triangle diagram cannot
explain the present peak.

All the above models lead to a fairly good description of the spectrum
and decay behavior of these two hidden-charm pentaquarks. To probe the underlying structures of these states, Wang et al.  \cite{Wang:2016dzu}
calculated their magnetic moments with $J^P={\frac{1}{2}}^{\pm}$,
${\frac{3}{2}}^{\pm}$, ${\frac{5}{2}}^{\pm}$, and ${\frac{7}{2}}^{+}$
in the molecular model, the diquark-triquark model, and the
diquark-diquark-antiquark model, respectively. In
fact, different color-flavor structures lead to very different
magnetic moments of the hidden-charm pentaquarks. In other words,
the magnetic moments may be a useful tool to distinguish various
models.

The axial charge of the baryon state is also a fundamental
observable. The axial charge of the nucleon was measured in the beta
decay of the polarized neutrons
\cite{Mendenhall:2012tz,Mund:2012fq}. The axial charge of the $P_c$
states can also be measured from their semileptonic decays in the
future. Moreover, since the width of the $P_c(4380)$ state is quite large,
its axial charge may be directly extracted from the process $P_c \to
P_c+\pi \to J/\psi+p+\pi$ once the $P_c(4380)$ state is produced.

The mass and magnetic moment of the hidden-charm
pentaquark state depend on both the light and the heavy quarks. In
contrast, the axial charge of the $P_c$ state is very sensitive to
the color, spin, and isospin configurations of the light quarks
inside the pentaquarks. Consequently, the axial charges provide a
new independent probe of the inner structures of the $P_c$ states.

The chiral symmetry and its spontaneous breaking play a pivotal role
in the low-energy regime, which inspired various versions of the
chiral quark model \cite{Liu:2008qb,
Zhang:1994pp,Manohar:1983md,Vijande:2004he,Valcarce:2005em}. The
long-range one-pion-exchange force is introduced to model the
interaction between the light quarks, together with the confinement
and the one-gluon-exchange potential
\cite{Barik:1986mq,Girdhar:2015gsa,Dahiya:2014jfa,Zhao:2010jc,Li:2013bca}.
The chiral quark models are widely applied to study the exotic
states. Some possible molecular states are predicted in Ref.
\cite{Li:2013bca} using the chiral SU(3) quark model. In Refs.
\cite{Yang:2015bmv,Ortega:2016syt}, the chiral quark model was used
to calculate the mass spectrum and decay width of the molecular
pentaquark states. In this work, we calculate the axial
charges of the pentaquark states.

This work is organized as follows. In Sec. \ref{sec2}, we construct
the spin-flavor wave functions of the $P_c$ states under the SU(2)
flavor symmetry. In Sec. \ref{sec3}, we use the chiral quark model
to derive the axial charges of the pentaquark states. In Sec.
\ref{sec4}, we study the transition coupling for the $P'_cP_c\pi$
interaction where $P'_c$ and $P_c$ are different. We illustrate why
the numerical values of the axial charges of the $P_c$ states are
generally smaller than that of the proton in Sec. \ref{sec5}. We
give a summary in Sec. \ref{sec6}. We also present the SU(3)
invariant strong interaction Lagrangians of the $P_c$ state and
pseudoscalar mesons in the Appendix.

%%%%%%%%%%%%%%%%%%%%%%%%%%%%%%%%%%%%%%%%%%%
\section{The spin-flavor wave functions of the $P_c$ states}\label{sec2}
%%%%%%%%%%%%%%%%%%%%%%%%%%%%

With the SU(2) flavor symmetry, we construct the wave functions of
the $P_c$ states as in Ref. \cite{Wang:2016dzu}. There are five
quarks, $\bar c c q_1q_2q_3$, in the $P_c$ states. In the molecular,
diquark-diquark-antiquark and diquark-triquark model, the $q_2q_3$
forms a quasibound diquark when they are in the antisymmetric $\bar
3_c$ color representation. The space wave function of the diquark is
symmetric. In the flavor space, $q_1q_2q_3$ may form the flavor
representations
\begin{eqnarray} \label{w0}
 2_{q_1}\otimes(2_{q_2}\otimes2_{q_3})=2_{q1}\otimes(1_f\oplus 3_f)=2_{1f}\oplus 2_{2f}\oplus
 4_f.
\end{eqnarray}
When the diquark $q_2q_3$ is in the antisymmetric $1_f$
representation, its isospin and spin are both $0$ due to the fermi
statistic, while the spin and isospin are both $1$ when the diquark
is in the symmetric $3_f$ representation. The isospin of the three
light quarks in the $2_{1f}$ or $2_{2f}$ representation is
$\frac{1}{2}$, while it is $\frac{3}{2}$ when they are in the $4_f$
representation. The explicit flavor wave functions of the three
light quarks in different representations are listed in Table
\ref{qqqflavor}. We can obtain the flavor wave functions of the
$P_c$ states by adding the heavy quark $c$ and antiquark $\bar c$
into the flavor wave functions of the three light quarks according
to the structures $(\bar Q q_1)(Q q_2 q_3)$, $(Q q_1)(\bar Q
q_2q_3)$, and $ (Q q_1)(q_2q_3)\bar Q$ in the molecular model, the
diquark-triquark model and the diquark-diquark-antiquark model, respectively.

\renewcommand{\arraystretch}{1.8}
\begin{table*}[htbp]
\caption{The flavor wave functions of the three light quarks
$q_2q_3q_1$ in the $P_c$ states. The $(I,I_3)$ denotes the isospin
and its third component, respectively. $\{q_2
q_3\}_{+}=\frac{1}{\sqrt 2}(q_2q_3+q_2q_3)$.
$[q_2q_3]_{-}=\frac{1}{\sqrt 2}(q_2q_3-q_3q_2)$. The two quarks in $[]_-$ and $\{\}_{+}$ form the quasibound diquark, while the
quark outside is from another cluster.}\label{qqqflavor}
\begin{center}
   \begin{tabular}{c|c|c} \toprule[1pt]\toprule[1pt]
 $( I, I_3)$ &\multicolumn{1}{c|} {Wave function $2_{1f}$}   &wave function $2_{2f}$\\
\midrule[1pt]

$(\frac{1}{2},\frac{1}{2})$ &$[ud]_{-}u$ &$\sqrt{\frac{2}{3}}\{uu\}_{+}d-\sqrt{\frac{1}{3}} \{ud\}_{+}u $\\
$(\frac{1}{2},-\frac{1}{2})$  &$[ud]_{-}d$ &$\sqrt{\frac{1}{3}}\{ud\}_{+}d-\sqrt{\frac{2}{3}}\{dd\}_{+}u$\\
\bottomrule[1pt]

$( I, I_3)$ &\multicolumn{1}{c|} {Wave function $4_f$}
&\\\midrule[1pt]
$(\frac{3}{2},\frac{3}{2})$& $\{uu\}_{+}u$     \\
$(\frac{3}{2},\frac{1}{2})$  &  $\sqrt{\frac{1}{3}}\{uu\}_{+}d+\sqrt{\frac{2}{3}}\{ud\}_{+}u$  \\
$(\frac{3}{2},-\frac{1}{2})$&  $\sqrt{\frac{2}{3}}\{ud\}_{+}d+\sqrt{\frac{1}{3}}\{dd\}_{+}u$  \\
$(\frac{3}{2},-\frac{3}{2})$&  $\{dd\}_{+}d$   \\
     \bottomrule[1pt]\bottomrule[1pt]
      \end{tabular}
  \end{center}
\end{table*}

Let us choose the construction of the flavor wave function of the
molecular pentaquark state $(\bar Q q_1)(Q q_2q_3)$ as an example.
As illustrated above, $q_2q_3$ may form a diquark in the
antisymmetric $\bar 3_c$ color representation with the spin and
isospin both being $0$. Now $(Q q_2q_3)$ is a color singlet baryon
corresponding to the baryon $\Lambda_Q$. $\bar Q q_1$ forms a
heavy meson. The $P_c$ state is a molecular state composed of $ \bar
D^{(\ast)}\Lambda_Q$ with $I=\frac{1}{2}$. When both the spin
and the isospin of the diquark are $1$, the pentaquark states are in the
$2_{2f}$ and $4_f$ representations with $I=\frac{1}{2}$ and $\frac{3}{2}$,
respectively. The molecular $P_c$ state is composed of $\bar
D^{(\ast)}\Sigma_c^{(\ast)}$. For example, the recently observed
$P_c(4380)$ state is supposed to be the $\bar
D^{(\ast)}\Sigma_c^{(\ast)}$states in the $2_{2f}$ representation
with $(I,I_3)=(\frac{1}{2},\frac{1}{2})$
\cite{Chen:2015moa,Chen:2015loa,Chen:2016qju,Shen:2016tzq,Lu:2016nnt}.
Their flavor wave functions are
\begin{eqnarray} \label{w1}
|P_c\rangle= \sqrt{\frac{2}{3}}(c\{uu\}_+)({\bar
cd})-\sqrt{\frac{1}{3}} (c\{ud\}_+)(\bar c u)=
\sqrt{\frac{2}{3}}\Sigma_c^{(\ast)++} \bar D^{(\ast)-}-\sqrt{\frac{1}{3}}
\Sigma_c^{(\ast)+} {\bar D}^{(\ast)0} .
\end{eqnarray}

After considering the spin wave functions of the $P_c$ states at the
same time, we obtain the spin-flavor wave functions. For example, if
the $P_c(4380)$ state is a $\frac{3}{2}^-$ molecular state composed of
$\bar D^{*}\Sigma_c$, its wave function reads
\begin{eqnarray} \label{w2}
&&|P_c^{\frac{3}{2},+\frac{3}{2}}\rangle=\Big \{ \sqrt{\frac{2}{3}}(c\{uu\}_+) ({\bar cd})-\sqrt{\frac{1}{3}} (c\{ud\}_+)(\bar c u)\Big \}\otimes \Big \{\sqrt{\frac{2}{3}}|\downarrow_Q\uparrow_{q_2}\uparrow_{q_3}\rangle|\uparrow_{\bar Q}\uparrow_{q_1}  \rangle-\sqrt{\frac{1}{3}}\frac{(|\uparrow_Q\downarrow_{q_2}\uparrow_{q_3}\rangle+|\uparrow_Q\uparrow_{q_2}\downarrow_{q_3}\rangle)|\uparrow_{\bar Q}\uparrow_{q_1}  \rangle}{\sqrt 2} \Big  \} \nonumber \\
 &&=\sqrt{\frac{2}{3}}|\Sigma_c^{++}, J_3=\frac{1}{2}\rangle |\bar D^{\ast-},J_3=1\rangle-\sqrt{\frac{1}{3}}| \Sigma_c^{+},J_3=\frac{1}{2} \rangle |{\bar D}^{\ast 0},J_3=1\rangle ,
\end{eqnarray}
where the superscripts represent the total angular momentum and its
third component. The up and down arrows denote the spin of the quark
up and down, respectively. The subscripts $Q$ and $q$ denote the
heavy quark and the light quark, respectively.

The wave functions of the other $P_c$ states with different isospins
are similar. We select the $\frac{3}{2}^{-}$ pentaquark state with
$(I,I_3)=(\frac{3}{2},\frac{1}{2})$ as an example,
\begin{eqnarray} \label{w001}
|P_c\rangle=\Big\{ \sqrt{\frac{1}{3}}(c\{uu\}_{+})(\bar c
d)+\sqrt{\frac{2}{3}}(\{ud\}_{+})(\bar c u) \Big \} \otimes \Big
\{\sqrt{\frac{2}{3}}|\downarrow_Q\uparrow_{q_2}\uparrow_{q_3}\rangle|\uparrow_{\bar
Q}\uparrow_{q_1}
\rangle-\sqrt{\frac{1}{3}}\frac{(|\uparrow_Q\downarrow_{q_2}\uparrow_{q_3}\rangle+|\uparrow_Q\uparrow_{q_2}\downarrow_{q_3}\rangle)|\uparrow_{\bar
Q}\uparrow_{q_1}  \rangle}{\sqrt 2} \Big  \} .
\end{eqnarray}

%%%%%%%%%%%%%%%%%%%%%%%%%%%%%%%%%
\section{The axial charge of the pentaquark state}
\label{sec3}
%%%%%%%%%%%%%%%%%%%%%%%%%%%%%%%

\subsection{The chiral quark model and the axial charge of the nucleon}\label{subsec1}

The Lagrangian of the chiral quark model is
\begin{eqnarray} \label{w7}
\mathscr{L}_{quark}=\frac{1}{2} g_1 \bar \psi_q\gamma^{\mu}\gamma_5
\partial_{\mu}\phi \psi_q\sim \frac{1}{2}g_1 \bar \psi_q \sigma_z
\partial_z\phi \psi_q = \frac{1}{2}\frac{g_1}{f_\pi} (\bar u
\sigma_z\partial_z\pi_0 u-\bar d \sigma_z\partial_z\pi_0 d)+...,
\end{eqnarray}
where $g_1$ is the coupling coefficient, and $f_\pi=92$ MeV is the decay constant of the pion. $\psi_q$($ \bar \psi_q$) is
the quark (antiquark) field. $u$ and $d$ are the up and down quarks,
respectively. In the calculation, the $z$ axis is along the momentum
of the meson $\pi$. $\sigma_z$ is the Pauli matrix for the light
quarks. $\phi$ represents the pseudoscalar meson field in the
SU(2) flavor symmetry:
\begin{equation} \label{w6}
\phi=\frac{1}{f_\pi}\left(
  \begin{array}{cc}
      \pi_0 &\sqrt{2}\pi^{+}\\
    \sqrt{2}\pi^{-} &-\pi_0\\
 \end{array}
\right).
\end{equation}
The $NN\pi_0$ Lagrangian is
 \begin{eqnarray} \label{w15}
 \mathscr{L}^{p}_{eff}=\frac{1}{2}g_{A} \bar N \gamma^{\mu} \gamma_5 \partial_{\mu}\phi N \sim \frac{g_{A}}{f_\pi}\bar N\frac{\Sigma_{Nz}}{2}\partial_z \pi_0 N,
 % g_{A}\bar N S_{\mu} \partial^{\mu}\pi_0 N=
 \end{eqnarray}
where $g_A$ is the axial charge of the nucleon.
$\frac{\Sigma_{Nz}}{2}$ is the $z$ component of the proton spin
operator.
\begin{eqnarray} \label{w170}
&&\langle  N, j_3=+\frac{1}{2}; ~\pi_0| \frac1i \frac{g_{A}}{f_\pi}\bar N\frac{\Sigma_{Nz}}{2}\partial_z \pi_0 N |N, j_3=+\frac{1}{2}\rangle = \frac{1}{2}\frac{q_z}{f_\pi}g_{A},\nonumber\\
\end{eqnarray}
where $q_z$ is the external momentum of $\pi_0$.
At the quark level,
\begin{eqnarray} \label{w16}
&&\langle  N , j_3=+\frac{1}{2} ;~\pi_0| \frac1i \frac{1}{2}\frac{g_1}{f_\pi} (\bar u
\sigma_z\partial_z\pi_0 u-\bar d \sigma_z\partial_z\pi_0 d)|N,
j_3=+\frac{1}{2}\rangle=\frac{5}{6}  \frac{q_z}{f_\pi}g_1.
\end{eqnarray}
From Eq. (\ref{w170}) and (\ref{w16}), we obtain
$g_1=\frac{3}{5}g_A$. In the next section, we use the same
formalism to derive the $P_cP_c\pi_0$ couplings in terms of $g_1$
and $g_A$.

\subsection{The axial charge of a $P_c$ state at the quark level}\label{subsec2}

We choose the $P_c$ states with $(I,I_3)=(\frac{1}{2},\frac{1}{2})$
in the $2_{2f}$ representation to illustrate the calculation. The
wave functions of these pentaquark states are
 \begin{eqnarray} \label{w8}
&&| JJ_z\rangle_A=\Big\{ \sqrt{\frac{2}{3}}(c\{uu\}_+) ({\bar cd})-\sqrt{\frac{1}{3}} (c\{ud\}_+)(\bar c u) \Big\} \otimes \Big| \Big[(c(q_2 q_3)_{s_{23}})_{s_t} \otimes {(\bar c q_1)_{s_{1Q}}} \Big]_{s}\otimes l \Big \rangle^{J_z} _J  ,  \nonumber\\
&&| JJ_z\rangle_B= \Big\{ \sqrt{\frac{2}{3}}(\bar c\{uu\}_+) ({ cd})-\sqrt{\frac{1}{3}} (\bar c\{ud\}_+)(c u)\Big\}\otimes\Big|\Big[(\bar c (q_2q_3)_{s_{23}})_{s_t}\otimes {( c q_1)_{s_{1Q}}} \Big]_{s}\otimes l \Big\rangle^{J_z} _J  ,  \nonumber\\
&&| JJ_z\rangle_C=\Big\{\sqrt{\frac{2}{3}}\bar c(\{uu\}_+)  ({ cd})-\sqrt{\frac{1}{3}} \bar c(\{ud\}_+)(c u) \Big \}  \otimes \Big| \Big[ \{  \bar c\otimes (q_2 q_3)_{s_{23}}\}_{s_t} \otimes{( c q_1)_{s_{1Q}}}]_{s}\otimes l \Big \rangle^{J_z} _J  ,  \nonumber\\
\end{eqnarray}
where we use the subscripts $A, B$ and $C$ to denote the states in the
molecular, diquark-triquark, and diquark-diquark-antiquark models,
respectively. $s_{23}$ is the spin of the diquark $q_2q_3$. $s_t$ is
the sum of $s_{23}$ and the spin of the heavy quark or antiquark.
$s_{1Q}$ is the spin of the diquark $c q_1$ or the meson $\bar c
q_1$, which couples with $s_t$ to form the total spin $s$. The sum
of $s$ and the orbital momentum $l$ leads to the total angular
momentum $J$. $J_z$ is its third component.

For the above $P_c$ state with $(I,I_3)=(\frac{1}{2},\frac{1}{2})$
in the molecular scheme,
\begin{eqnarray} \label{w9}
&&\langle P^{J,J}_c\pi_0|\frac1i\frac{1}{2}\frac{g_1}{f_\pi} (\bar u
\sigma_z\partial_z\pi_0 u-\bar d \sigma_z\partial_z\pi_0 d)| P'^{J',J}_c \rangle=\frac{g_1}{f_\pi}\Big\{\frac{2}{3}\langle (c\{uu\}_+) ({\bar cd})|(\bar u u-\bar dd)|(c\{uu\}_+) ({\bar cd})\rangle\nonumber \\
&&+\frac{1}{3}\langle (c\{ud\}_+)(\bar c u) |(\bar u u-\bar dd)|  (c\{ud\}_+)(\bar c u)  \rangle \Big\}\otimes \langle \Big[ [{(\bar c q_1)_{s_{1Q}}}\otimes(c\otimes (q_2 q_3)_{s_{23}})_{s_t} ]_{s}\otimes l\Big ]^J _J | \frac{1}{2}\sigma_z q_z|\Big [[{(\bar c q_1)_{{s'_{1Q}}}}\otimes(c\otimes (q_2 q_3)_{s'_{23}})_{s'_t} ]_{s'}\otimes l'\Big]_{J'} ^ J\rangle  \nonumber\\
&&= \sum_{m,s_z,m',s'_z} \sum_{s_{{1Q}_z}s'_{{1Q}_z},s_{t_z},s'_{t_z}}  \sum_{s_{\bar Q_z}s_{1_z},s'_{\bar Q_z}s'_{1_z}}  \sum_{s_{Q_z},s_{23_z},s'_{Qz},s'_{23_z}} \sum_{s_{2_z},s_{3_z},s'_{2_z},s'_{3_z}} \langle ss_z, lm|J J\rangle^{\ast}\langle s's'_z, l'm'|J' J\rangle \langle  s_t s_{t_z}, s_{1Q} s_{1Q_z}|ss_z\rangle^{\ast}\langle s'_t s'_{t_z}, s'_{1Q} s'_{1Q_z}|s' s'_z\rangle \nonumber\\
&& \langle \frac{1}{2}s_{{\bar Q}_z},\frac{1}{2}s_{1_z}|s_{1Q}s_{1Q_z}\rangle^{\ast} \langle \frac{1}{2}s'_{{\bar Q}_z},\frac{1}{2}s_{1_z}|s'_{1Q}s'_{1Q_z}\rangle \langle \frac{1}{2}s_{Q_z},s_{23} s_{23_z}|s_ts_{t_z}\rangle^{\ast}\langle \frac{1}{2}s'_{Q_z},s'_{23} s'_{23_z}|s'_ts'_{t_z}\rangle \langle \frac{1}{2} s_{2_z} ,\frac{1}{2} s_{3_z}|s_{23}s_{23_z}\rangle^{\ast} \langle \frac{1}{2} s'_{2_z}, \frac{1}{2} s'_{3_z}|s'_{23}s'_{23_z}\rangle \nonumber\\
&&\frac{g_1}{f_\pi}\langle s_{1_z}s_{2_z}s_{3_z}s_{Q_z}s_{{\bar Q}_z}|(-\frac{1}{3}s_{1_z}+\frac{2}{3}s_{2_z}+\frac{2}{3}s_{3_z})| s'_{1_z}s'_{2_z}s'_{3_z}s'_{Q_z}s'_{{\bar Q}_z}\rangle q_z\langle P_c(lm)|P'_c(l'm') \rangle \nonumber\\
&&= \sum_{m,s_z;s_{1Q_z},s_{t_z}}  \sum_{s_{\bar Q_z}s_{1_z};
s_{Q_z},s_{23_z}} \sum_{s_{2_z},s_{3_z}}
g_1W_{s,s_z,l,m,s_{1Q},s_{t_z},s',l'}^{s_{1_z},s_{2_z},s_{3_z},s_{Qz},s_{\bar
Q_z}}
(-\frac{1}{3}s_{1_z}+\frac{2}{3}s_{2_z}+\frac{2}{3}s_{3_z}) \frac{q_z}{f_\pi}\langle  P_c(lm)|P'_c(l'm') \rangle \nonumber\\
&&= \sum_{m,s_z;s_{1Q_z},s_{t_z}}  \sum_{s_{\bar Q_z}s_{1_z};
s_{Q_z},s_{23_z}} \sum_{s_{2_z},s_{3_z}}
g_1W_{s,s_z,l,m,s_{1Q},s_{t_z},s',l'}^{s_{1_z},s_{2_z},s_{3_z},s_{Qz},s_{\bar
Q_z}}O_{123}\frac{q_z}{f_\pi},
\end{eqnarray}
where $J\leqslant J'$. $W$ denotes the product of all the
Clebsch-Gordan coefficients. $O_{123}$ is the spin operator. Its
expression is given in Eqs. (\ref{w01})-(\ref{w04}). The other
notations are the same as those in Eq. (\ref{w8}). $P_c(lm)$ is the
space wave function of the $P_c$ state with the orbital quantum
number $(l,m)$. In the last line, we have used the heavy quark
symmetry. The $\bar D$ and $\bar D^{*}$, $\Sigma_c$ and $\Sigma^*_c$
form the doublets respectively. Their space wave functions are the
same only if the $P_c$ states have the same $(l,m)$ in the heavy
quark limit. Their space wave functions with different $(l,m)$ are
orthogonal to each other. In the transition process, the space
overlapping amplitude does not depend on $m$. For example, in the
molecular model, the $s$-wave $P_c$ states composed of
$D^{(*)}\Sigma_c^{(*)}$ with the same isospin belong to one
multiplet. Their space wave functions are the same. After exchanging
the heavy quark and antiquark in Eq. (\ref{w9}), we obtain the
equations for the diquark-triquark and diquark-diquark-antiquark
states. For the pentaquark states with the other isospin factors,
the $O_{123}$ is different due to the different flavor wave
functions in Eq. (\ref{w9}). For the $P_c$ state in the $2_{1f}$
representation with $(I,I_3)=(\frac{1}{2},\pm \frac{1}{2})$,
\begin{eqnarray} \label{w01}
O_{123}=\pm s_{1_z}.
\end{eqnarray}
For the $P_c$ state in the $2_{2f}$ representation with
$(I,I_3)=(\frac{1}{2},\pm \frac{1}{2})$,
\begin{eqnarray} \label{w02}
O_{123}=\pm (-\frac{1}{3}s_{1_z}+\frac{2}{3}s_{2_z}+\frac{2}{3}s_{3_z}).
\end{eqnarray}
For the $P_c$ state in the $4_{f}$ representation with
$(I,I_3)=(\frac{3}{2},\pm \frac{3}{2})$,
\begin{eqnarray} \label{w03}
O_{123}=\pm (s_{1_z}+s_{2_z}+s_{3_z}).
\end{eqnarray}
For the $P_c$ state in the $4_{f}$ representation with
$(I,I_3)=(\frac{3}{2},\pm \frac{1}{2})$,
\begin{eqnarray} \label{w04}
&&O_{123}^{I_3=\frac{1}{2}}=\pm\frac{1}{3}(s_{1_z}+s_{2_z}+s_{3_z}).
\end{eqnarray}
%We note that the $q_2q_3$ pair is symmetric when the $P_c$ state is
%in the $4_f$ representation. Thus, for the pentaquark states with
%$(I,I_3)=(\frac{3}{2},\pm \frac{1}{2})$, $O_{123}^{I_3=\pm
%\frac{1}{2}}=\pm (\frac{1}{3}s_{1_z}+s_{2_z}-\frac{1}{3}s_{3_z})$.

In Tables \ref{rd}, \ref{rdcs}, and \ref{rdc}, we list the results
for the $P_c$ states in the $2_{2f}$ flavor representation with
$(I,I_3)=(\frac{1}{2},\frac{1}{2})$ in the molecular scheme. We can
also obtain the analytical expressions for the states with the
quantum numbers in the other two models through exchanging $c$ and
$\bar c$ in Eq. (\ref{w9}). We note that the pionic interactions
only exist between the light quarks and do not depend on the heavy
quark and the antiquark. If the corresponding states in the three
different models have the same inner angular quantum numbers such as $s_{23}$, $s_t$, etc., their axial charges are the same. For
instance, in Table \ref{rd}, the $\frac{1}{2}^-$ $P_c$ state with
the configuration $|^2S_{\frac{1}{2}},0^{-}\otimes
\frac{1}{2}^{+}\otimes 0^{+} \rangle$ represents an $s$-wave molecular
state composed of $D\Sigma_c$ in the molecular model. In the
diquark-triquark model, the similar configuration
$|^2S_{\frac{1}{2}},0^{+}\otimes \frac{1}{2}^{-}\otimes 0^{+}
\rangle$ denotes the $\frac{1}{2}^-$ diquark-triquark state
$|(cq_1)_{0}(\bar c(q_2q_3)_1)_{\frac{1}{2}}\rangle$. In the
diquark-diquark-antiquark model, $|^2S_{\frac{1}{2}},0^{+}\otimes
\frac{1}{2}^{-}\otimes 0^{+} \rangle$ corresponds to the state
$|(cq_1)_{0}\otimes \{\bar c\otimes
(q_2q_3)_1\}_{\frac{1}{2}}\rangle$. All these three states lead to
the same axial charge. Similarly, the other results
in Table \ref{rd} can also be interpreted as the axial charges of
the corresponding states in the other two models.

When the $P_c$ states are in the $2_{1f}$ flavor representation, the molecular
pentaquark states are composed of $\bar D^{(*)} \Lambda_c$. However, there does not exist the pionic interaction between $\bar D^{(*)}$ and $\Lambda_c$ in the $\bar D^{(*)} \Lambda_c$ system. Thus, we list the results for the pentaquark states in
the diquark-triquark scheme in Table \ref{dtrd}. If we decompose the
$\frac{1}{2}^{-}$ triquark in the third column into {$\frac{1}{2}^{-}_{\bar
c}\otimes 0^{+}_{(q_2q_3)}$}, we obtain the axial charges
in the diquark-diquark-antiquark scheme.

\subsection{The axial charge of a $P_c$ state at the hadron level}\label{subsec3}

For the $P_c$ states with the isospin $\frac{1}{2}$, the $P_c P_c
\pi$ Lagrangians at the hadron level are
 \begin{eqnarray} \label{w11}
 &&\mathscr{L}^{\frac{1}{2}}_{eff}=\frac{1}{2}g_{P_1} \bar P_c \gamma^{\mu} \gamma_5 \partial_{\mu}\phi P_{c} \sim \frac{g_{P_1}}{f_\pi}\bar P_c\frac{\Sigma_z}{2}\partial_z \pi_0 P_{c},\nonumber \\
&&\mathscr{L}^{\frac{3}{2}}_{eff}=\frac{1}{2} g_{P_3}\bar P_c^{\nu}\gamma^{\mu} \gamma_5 \partial_{\mu}\phi P_{c\nu}\sim \frac{g_{P_3}}{f_\pi}\bar P_c^{\nu}\frac{\Sigma_z}{2}\partial_z \pi_0 P_{c\nu},\nonumber \\
&&\mathscr{L}^{\frac{5}{2}}_{eff}=\frac{1}{2} g_{P_5}\bar
P_c^{\alpha \beta}\gamma^{\mu} \gamma_5 \partial_{\mu}\phi P_{c
{\alpha \beta}}\sim \frac{g_{P_5}}{f_\pi}\bar P_c^{\alpha \beta}\frac{\Sigma_z}{2}
\partial_z \pi_0 P_{c \alpha \beta},
\end{eqnarray}
where $P_c$, $P_{c\nu}$, and $P_c^{ \alpha \beta} $ are the
$\frac{1}{2}^{\pm}$, $\frac{3}{2}^{\pm}$, and $\frac{5}{2}^{\pm}$
pentaquark states, respectively. $g_{P_{1(3,5)}}$ are the coupling
coefficients. $\frac{\Sigma_{z}}{2}$ is the spin operator of the
$P_c$ states.

For the $P_c$ state with $(I,I_3)=(\frac{1}{2},\frac{1}{2})$,
\begin{eqnarray} \label{w13}
&&\langle P_c^{\frac{1}{2},+\frac{1}{2}} \pi_0|\frac1i\frac{g_{P_1}}{f_\pi}\bar P_c\frac{\Sigma_z}{2}\partial_z \pi_0 P_{c}|P_c^{\frac{1}{2},+\frac{1}{2}}   \rangle = \frac{1}{2}\frac{q_z}{f_\pi}g_{P_1},\nonumber\\
&&\langle P_c^{\frac{3}{2},+\frac{3}{2}} \pi_0|\frac1i \frac{g_{P_3}}{f_\pi}\bar P_c^{\nu}\frac{\Sigma_z}{2}\partial_z \pi_0 P_{c\nu}|P_c^{\frac{3}{2},+\frac{3}{2}}   \rangle = \frac{3}{2}\frac{q_z}{f_\pi}g_{P_3},\nonumber\\
&&\langle P_c^{\frac{5}{2},+\frac{5}{2}}\pi_0 |\frac1i   \frac{g_{P_5}}{f_\pi}\bar P_c^{\alpha \beta}\frac{\Sigma_z}{2}
\partial_z \pi_0 P_{c \alpha \beta} |P_c^{\frac{5}{2},+\frac{5}{2}}   \rangle
=\frac{5}{2}\frac{q_z}{f_\pi}g_{P_5},
% \propto  g_{P_c}\sum_{m,s_z} s_z| \langle ss_z,lm|\frac{5}{2},\frac{5}{2}\rangle|^2
\end{eqnarray}
where the superscripts represent the total spin and its third
component, respectively. When the $P_c$ states are in the $2_{2f}$
and $2_{1f}$ representations, we list the  $P_cP_c\pi_0$ coupling
constants in Tables \ref{rd} and \ref{dtrd}, respectively.

 \renewcommand{\arraystretch}{1.8}
\begin{table*}[htbp]
\caption{The axial charges of the molecular pentaquark states at the
quark level and the hadron level. The $P_c$ states are in the
$2_{2f}$ representation with $(I,I_3)=(\frac{1}{2},\frac{1}{2})$.
The $P_c$ states are denoted as $|P_c^{J,J_3}\rangle$. In the third
column, we denote the angular momentum coupling according to meson
$\otimes$ baryon $\otimes$ orbital excitation. We also list the
constituent meson and baryon in the parentheses. }\label{rd}
\begin{center}
   \begin{tabular}{c|c|c|c|c|c|c} \toprule[1pt]\toprule[1pt]

 $J^P$ & & Molecular state&$\langle P^{\frac{1}{2},\frac{1}{2}}_c  \pi_0 |\frac{1}{i}\mathscr{L}_{quark}| P^{\frac{1}{2},\frac{1}{2}}_c  \rangle$/$g_1$  &$\langle P^{\frac{1}{2},+\frac{1}{2}}_c \pi_0 |\frac{1}{i}\mathscr{L}_{eff}| P^{+\frac{1}{2},\frac{1}{2}}_c  \rangle$/$g_{P_1}$ & $g_{P_1}$/$g_A$ & $g_{P_1}$\\
\hline
$\frac{1}{2}^{-}$  &$^2S_{\frac{1}{2}}$ & $0^{-}\otimes\frac{1}{2}^{+}\otimes 0^{+}(\bar D\Sigma_c)$  &$\frac{4}{9}$ &$\frac{1}{2}$ &$\frac{8}{15}$ &0.68\\

  & & $1^{-}\otimes\frac{1}{2}^{+}\otimes 0^{+}(\bar D^{\ast}\Sigma_c)$  &$-\frac{7}{27}$ &$\frac{1}{2}$ &-$\frac{14}{45}$ &-0.39\\
   && $1^{-}\otimes\frac{3}{2}^{+}\otimes 0^{+}(\bar D^{\ast}\Sigma_c^{\ast})$  &$\frac{23}{54}$ &$\frac{1}{2}$ &$\frac{23}{45}$ &0.65\\

   \hline
 $\frac{1}{2}^{+}$ &$^2P_{\frac{1}{2}}$ & $0^{-}\otimes\frac{1}{2}^{+}\otimes 1^{-}(\bar D\Sigma_c)$  &$-\frac{4}{27}$ &$\frac{1}{2}$ &$-\frac{8}{45}$ &-0.23\\
&$^2P_{\frac{1}{2}}$& ($1^{-}\otimes\frac{1}{2}^{+})_{\frac{1}{2}}\otimes1^{-}(\bar D^{\ast}\Sigma_c)$  &$\frac{7}{81}$ &$\frac{1}{2}$  &$\frac{14}{135}$ &0.13\\
&$^4P_{\frac{1}{2}}$ & ($1^{-}\otimes\frac{1}{2}^{+})_{\frac{3}{2}}\otimes1^{-}(\bar D^{\ast}\Sigma_c)$  &$\frac{25}{162}$ &$\frac{1}{2}$ &$\frac{5}{27}$ &0.23\\
&$^4P_{\frac{1}{2}}$ & ($0^{-}\otimes\frac{3}{2}^{+})_{\frac{3}{2}}\otimes1^{-}(\bar D\Sigma_c^{\ast})$  &$\frac{10}{27}$ &$\frac{1}{2}$ &$\frac{4}{9}$ &0.56\\
%& ($1^{-}\otimes\frac{3}{2}^{+})_{\frac{1}{2}}\otimes1^{-}(D^{\ast}\Sigma_c^{\ast})$  &-$\frac{23{54}$ &$\frac{1}{2}$\\
&$^2P_{\frac{1}{2}}$ & ($1^{-}\otimes\frac{3}{2}^{+})_{\frac{1}{2}}\otimes1^{-}(\bar D^{\ast}\Sigma_c^{\ast})$  &$-\frac{23}{162}$ &$\frac{1}{2}$ & $-\frac{23}{135}$ &-0.22\\
&$^4P_{\frac{1}{2}}$& ($1^{-}\otimes\frac{3}{2}^{+})_{\frac{3}{2}}\otimes1^{-}(\bar D^{\ast}\Sigma_c^{\ast})$  &$\frac{19}{81}$ &$\frac{1}{2}$ &$\frac{38}{135}$ &0.36\\
\hline

 $J^P$ & & Molecular state&$\langle P^{\frac{3}{2},\frac{3}{2}}_c  \pi_0|\frac{1}{i}\mathscr{L}_{quark}| P^{\frac{3}{2},+\frac{3}{2}}_c  \rangle$/$g_1$  &$\langle P^{\frac{3}{2},+\frac{3}{2}}_c  \pi_0|\frac{1}{i}\mathscr{L}_{eff}| P^{\frac{3}{2},\frac{3}{2}}_c \rangle$/$g_{P_3}$ & $g_{P_3}$/$g_A$ & $g_{P_3}$\\
\hline
$\frac{3}{2}^{-}$  &$^4S_{\frac{3}{2}}$ & $0^{-}\otimes\frac{3}{2}^{+}\otimes 0^{+}(\bar D\Sigma_c^{\ast})$  &$\frac{2}{3}$ &$\frac{3}{2}$ &$\frac{4}{15}$ &0.34\\
& &$1^{-}\otimes\frac{1}{2}^{+}\otimes 0^{+}(\bar D^{\ast}\Sigma_c)$  &$\frac{5}{18}$ &$\frac{3}{2}$ &$\frac{1}{9}$ &0.14\\
& &$1^{-}\otimes\frac{3}{2}^{+}\otimes 0^{+}(\bar D^{\ast}\Sigma_c^{\ast})$  &$\frac{19}{45}$ &$\frac{3}{2}$ &$\frac{38}{225}$ &0.21\\
\hline
$\frac{3}{2}^{+}$ &$^2P_{\frac{3}{2}}$ & $0^{-}\otimes\frac{1}{2}^{+}\otimes 1^{-}(\bar D\Sigma_c)$  &$\frac{4}{9}$ &$\frac{3}{2}$ &$\frac{8}{45}$ &0.23\\
&$^2P_{\frac{3}{2}}$& ($1^{-}\otimes\frac{1}{2}^{+})_{\frac{1}{2}}\otimes1^{-}(\bar D^{\ast}\Sigma_c)$  &-$\frac{7}{27}$ &$\frac{3}{2}$ &$-\frac{14}{135}$ &-0.13\\
&$^4P_{\frac{3}{2}}$ & ($1^{-}\otimes\frac{1}{2}^{+})_{\frac{3}{2}}\otimes1^{-}(\bar D^{\ast}\Sigma_c)$  &$\frac{11}{54}$ &$\frac{3}{2}$ &$\frac{11}{135}$ &0.10\\
&$^4P_{\frac{3}{2}}$ & ($0^{-}\otimes\frac{3}{2}^{+})_{\frac{3}{2}}\otimes1^{-}(\bar D\Sigma_c^{\ast})$  &$\frac{22}{45}$ &$\frac{3}{2}$ &$\frac{44}{225}$ &0.25\\
%& ($1^{-}\otimes\frac{3}{2}^{+})_{\frac{1}{2}}\otimes1^{-}(D^{\ast}\Sigma_c^{\ast})$  &-$\frac{23{54}$ &$\frac{1}{2}$\\
&$^2P_{\frac{3}{2}}$ & ($1^{-}\otimes\frac{3}{2}^{+})_{\frac{1}{2}}\otimes1^{-}(\bar D^{\ast}\Sigma_c^{\ast})$  &$\frac{23}{54}$ &$\frac{3}{2}$ & $\frac{23}{135}$ &0.22\\
&$^4P_{\frac{3}{2}}$& ($1^{-}\otimes\frac{3}{2}^{+})_{\frac{3}{2}}\otimes1^{-}(\bar D^{\ast}\Sigma_c^{\ast})$  &$\frac{209}{675}$ &$\frac{3}{2}$&$\frac{418}{3375}$ &0.15\\
&$^6P_{\frac{3}{2}}$ & ($1^{-}\otimes\frac{3}{2}^{+})_{\frac{5}{2}}\otimes1^{-}(\bar D^{\ast}\Sigma_c^{\ast})$  &$\frac{21}{50}$ &$\frac{3}{2}$  &$\frac{21}{125}$ & 0.22\\
\hline
$J^P$& & Molecular state &$\langle P^{\frac{5}{2},\frac{5}{2}}_c \pi_0 |\frac{1}{i}\mathscr{L}_{quark}| P^{\frac{5}{2},\frac{5}{2}}_c  \rangle$/$g_1$  &$\langle P^{\frac{5}{2},+\frac{5}{2}}_c \pi_0 |\frac{1}{i}\mathscr{L}_{eff}| P^{\frac{5}{2},+\frac{5}{2}}_c \rangle$/$g_{P_5}$ & $g_{P_5}$/$g_A$ & $g_{P_5}$\\
\hline
$\frac{5}{2}^{-}$  &$^6S_{\frac{5}{2}}$ &$1^{-}\otimes\frac{3}{2}^{+} \otimes0^{+}(\bar D^{\ast}\Sigma_c^{\ast})$  &$\frac{1}{2}$ &$\frac{5}{2}$&$\frac{3}{25}$ & 0.15\\
\hline
$\frac{5}{2}^{+}$ &$^4P_{\frac{5}{2}}$  & $1^{-}\otimes\frac{1}{2}^{+}\otimes 1^{-}(\bar D^{\ast}\Sigma_c)$  &$\frac{5}{18}$ &$\frac{5}{2}$&$\frac{1}{15}$ &0.08\\
  &$^4P_{\frac{5}{2}}$& $0^{-}\otimes \frac{3}{2}^{+}\otimes 1^{-}(\bar D\Sigma_c^{\ast})$  &$\frac{2}{3}$ &$\frac{5}{2}$ &$\frac{4}{25}$ &0.20\\
&$^4P_{\frac{5}{2}}$ & ($1^{-}\otimes\frac{3}{2}^{+})_{\frac{3}{2}}\otimes1^{-}(\bar D^{\ast}\Sigma_c^{\ast})$  &$\frac{19}{45}$ &$\frac{5}{2}$ &$\frac{38}{375}$ &0.13\\
&$^6P_{\frac{5}{2}}$ & ($1^{-}\otimes\frac{3}{2}^{+})_{\frac{5}{2}}\otimes1^{-}(\bar D^{\ast}\Sigma_c^{\ast})$  &$\frac{31}{70}$ &$\frac{5}{2}$&$\frac{93}{875}$ & 0.13\\
 \bottomrule[1pt]\bottomrule[1pt]
      \end{tabular}
  \end{center}
\end{table*}

 \renewcommand{\arraystretch}{1.8}
\begin{table*}[htbp]
\caption{The axial charges of the pentaquark when they belong to the
$2_{1f}$ representation with $(I,I_3)=(\frac{1}{2},\frac{1}{2})$. In
the third column, we use $cq_1\otimes ( \bar c(q_2q_3)_0) \otimes$
orbital excitation to denote the angular momentum
configurations.}\label{dtrd}
\begin{center}
   \begin{tabular}{c|c|c|c|c|c|c} \toprule[1pt]\toprule[1pt]

 $J^P$ & & &$\langle P^{\frac{1}{2},+\frac{1}{2}}_c  \pi_0|\frac{1}{i}\mathscr{L}_{quark}| P^{+\frac{1}{2},\frac{1}{2}}_c \rangle$/$g_1$  &$\langle P^{\frac{1}{2},+\frac{1}{2}}_c \pi_0 |\frac{1}{i}\mathscr{L}_{eff}| P^{+\frac{1}{2},\frac{1}{2}}_c  \rangle$/$g_{P_1}$ & $g_{P_1}$/$g_A$ & $g_{P_1}$\\
\hline
$\frac{1}{2}^{-}$  &$^2S_{\frac{1}{2}}$ & $0^{+}\otimes\frac{1}{2}^{-}\otimes 0^{+}$  &$0$ &$\frac{1}{2}$ &$0$ &$0$\\

  & & $1^{+}\otimes\frac{1}{2}^{-}\otimes 0^{+}$  &$\frac{1}{3}$ &$\frac{1}{2}$ &$\frac{2}{5}$ &0.51\\
     \hline
 $\frac{1}{2}^{+}$ &$^2P_{\frac{1}{2}}$ & $0^{+}\otimes\frac{1}{2}^{-}\otimes 1^{-}$  &$0$ &$\frac{1}{2}$ &$0$ &0\\
&$^2P_{\frac{1}{2}}$& ($1^{+}\otimes\frac{1}{2}^{-})_{\frac{1}{2}}\otimes1^{-}$  &-$\frac{1}{9}$ &$\frac{1}{2}$  &$-\frac{2}{15}$ &-0.17\\
&$^4P_{\frac{1}{2}}$ & ($1^{+}\otimes\frac{1}{2}^{-})_{\frac{3}{2}}\otimes1^{-}$  &$\frac{5}{18}$ &$\frac{1}{2}$ &$\frac{1}{3}$ &0.42\\
\hline

 $J^P$ & &&$\langle P^{\frac{1}{2},+\frac{3}{2}}_c \pi_0 |\frac{1}{i}\mathscr{L}_{quark}| P^{+\frac{1}{2},\frac{3}{2}}_c  \rangle$/$g_1$  &$\langle P^{\frac{1}{2},+\frac{3}{2}}_c \pi_0 |\frac{1}{i}\mathscr{L}_{eff}| P^{+\frac{1}{2},\frac{3}{2}}_c \rangle$/$g_{P_3}$ & $g_{P_3}$/$g_A$ & $g_{P_3}$\\
\hline
$\frac{3}{2}^{-}$ & &$1^{+}\otimes\frac{1}{2}^{-}\otimes 0^{+}$  &$\frac{1}{2}$ &$\frac{3}{2}$ &$\frac{1}{5}$ &0.25\\
\hline
$\frac{3}{2}^{+}$ &$^2P_{\frac{3}{2}}$ & $0^{+}\otimes\frac{1}{2}^{-}\otimes 1^{-}$  &$0$ &$\frac{3}{2}$ &$0$ &0\\
&$^2P_{\frac{3}{2}}$& ($1^{+}\otimes\frac{1}{2}^{-})_{\frac{1}{2}}\otimes1^{-}(D^{\ast}\Sigma_c)$  &$\frac{1}{3}$ &$\frac{3}{2}$ &$\frac{2}{15}$ &0.17\\
&$^4P_{\frac{3}{2}}$ & ($1^{+}\otimes\frac{1}{2}^{-})_{\frac{3}{2}}\otimes1^{-}(D^{\ast}\Sigma_c)$  &$\frac{11}{30}$ &$\frac{3}{2}$ &$\frac{11}{75}$ &0.19\\
\hline
$J^P$& & &$\langle P^{\frac{1}{2},+\frac{5}{2}}_c  \pi_0 |\frac{1}{i}\mathscr{L}_{quark}| P^{+\frac{1}{2},\frac{5}{2}}_c  \rangle$/$g_1$  &$\langle P^{\frac{1}{2},+\frac{5}{2}}_c  \pi_0|\frac{1}{i}\mathscr{L}_{eff}| P^{+\frac{1}{2},\frac{5}{2}}_c \rangle$/$g_{P_5}$ & $g_{P_5}$/$g_A$ & $g_{P_5}$\\
\hline

$\frac{5}{2}^{+}$ &$^4P_{\frac{5}{2}}$  & $1^{+}\otimes\frac{1}{2}^{-}\otimes 1^{-}$  &$\frac{1}{2}$ &$\frac{5}{2}$&$\frac{3}{25}$ &0.15\\
 \bottomrule[1pt]\bottomrule[1pt]
      \end{tabular}
  \end{center}
\end{table*}

The $P_c$ states with the same angular momentum may have very
different axial charges if their parities and internal
configurations are different. For example, the
$P_{c}(^{4}S_{\frac{3}{2}})$ states may be composed of $\bar
D\Sigma_c^{\ast}$ or $\bar D^{\ast}\Sigma_c$, while their axial charges are $0.34$ and $0.14$, respectively. The
$\frac{1}{2}^+$ $P_c$ state may be composed of $\bar
D^{\ast}\Sigma_c^{\ast}$. The spin of $\bar
D^{\ast}\Sigma_c^{\ast}$ is either $\frac{1}{2}$ or $\frac{3}{2}$.
The pentaquark states have the corresponding configurations
$^2P_{\frac{1}{2}}$ and $^4P_{\frac{1}{2}}$, respectively. Their
axial charges are $-0.22$ and $0.36$, respectively.

%%%%%%%%%%%%%%%%%%%%%%%%%%%%%%%%%%%%
\section{The transition pionic coupling for the $P'_cP_c\pi$ interaction}\label{sec4}
%%%%%%%%%%%%%%%%%%%%%%%%%%%%%%%%%%%%

In this section, we study the transition pionic couplings between
two different pentaquark states when they have the same parity. For the
process $P'_c(J^P=\frac{1}{2}^{\pm})\rightarrow
P_c(J^P=\frac{1}{2}^{\pm}) \pi$,
$P'_c(J^P=\frac{3}{2}^{\pm})\rightarrow P_c(J^P=\frac{3}{2}^{\pm})
\pi$ and $P'_c(J^P=\frac{5}{2}^{\pm})\rightarrow
P_c(J^P=\frac{5}{2}^{\pm}) \pi$, the Lagrangian is the same as that in Eq.
(\ref{w11}) with $ P_c$ replaced by $ P'_c$. We use $g_{11}$,
$g_{33}$, and $g_{55}$ to denote the $P'_cP_c\pi_0$ coupling when
the parities of $P'_c$ and $P_c$ are negative. We use $g'_{11}$,
$g'_{33}$, and  $g'_{55}$ to denote the $P'_cP_c\pi_0$ coupling when
the parities of $P'_c$ and $P_c$ are positive.

In Table \ref{rdcs}, we list the results. When the two pentaquark
states have the same $J^P$ but different configurations, these
transition pionic couplings are smaller than $0.19$ except that
for $P'_c(\bar D^{\ast}\Sigma_c,^4P_{\frac{1}{2}})P_c(\bar
D^{\ast}\Sigma_c,^2P_{\frac{1}{2}})\pi_0$. Some of them are
tiny, which indicates the interaction between the $P_c$ states with
the same $J^P$ but different configurations is very weak. Some
pentaquark states can not interact with each other through the
pseudoscalar field and their couplings vanish. The reasons are as
follows. For the states $\bar D\Sigma_c$ and $\bar
D^{\ast}\Sigma_c^{\ast}$, there should be at least two spin
operators at the quark level to flip the spin of both the heavy
meson and the baryon. However, there is only one spin operator
$\sigma_z$ in the Lagrangian in Eq. (\ref{w7}). Thus, their
couplings are zero. The same conclusion holds for the $\bar
D^{\ast}\Sigma_c$ and $\bar D\Sigma_c^{\ast}$ transitions. Moreover,
the single spin operator $\sigma_z$ cannot mix the states
$P'_c(^2L_J)$ and $P_c(^6L_J)$ through the pion exchange.

\renewcommand{\arraystretch}{1.6}
\begin{table*}[htbp]
\caption{The transition coupling constants of the molecular
pentaquark states at the quark level and the hadron level. The two
pentaquark states are in the $2_{2f}$ representation with
$(I,I_3)=(\frac{1}{2},\frac{1}{2})$. The $P'_c$ and $P_c$ states have the
same $J^P$ quantum numbers but different configurations. The
superscripts denote the total angular momentum and its third
component. }\label{rdcs}
\begin{center}
   \begin{tabular}{c|c|c|c|c|c} \toprule[1pt]\toprule[1pt]
 &&$\langle P^{\frac{1}{2},\frac{1}{2}}_c \pi_0 |\frac1i\mathscr{L}_{quark}| P'^{\frac{1}{2},\frac{1}{2}}_c \rangle$/$g_1$  &$\langle P^{\frac{1}{2},\frac{1}{2}}_c \pi_0|\frac{1}{i}\mathscr{L}_{eff}| P'^{\frac{1}{2},\frac{1}{2}}_c \rangle$/$g_{22}$ & $g_{22}$/$g_A$&  $g_{22}$\\
\hline

$\frac{1}{2}^+$ &$\langle \bar D \Sigma_c, ^2P_{\frac{1}{2}}|\mathscr{L}|  \bar D^{\ast}\Sigma_c, ^2P_{\frac{1}{2}}\rangle$ & $-\frac{1}{18\sqrt 3}$ &$\frac{1}{2}$ & $-\frac{1}{15 \sqrt 3}$ &$-0.05$\\
 &$\langle \bar D \Sigma_c, ^2P_{\frac{1}{2}}|\mathscr{L}|\bar  D^{\ast}\Sigma_c, ^4P_{\frac{1}{2}}\rangle$ & $-\frac{1}{9}\sqrt{\frac{2}{3}}$ &$\frac{1}{2}$ & $-\frac{2}{15}\sqrt{\frac{2}{3}}$&$-0.14$\\
&$\langle\bar D^{\ast} \Sigma_c, ^4P_{\frac{1}{2}}|\mathscr{L}|\bar  D^{\ast}\Sigma_c, ^2P_{\frac{1}{2}}\rangle$ & $-\frac{19\sqrt 2}{81}$ &$\frac{1}{2}$ & $-\frac{38\sqrt 2}{135}$ &$-0.50$\\

&$\langle\bar D^{\ast} \Sigma_c^{\ast}, ^2P_{\frac{1}{2}}|\mathscr{L}|\bar  D\Sigma_c, ^2P_{\frac{1}{2}}\rangle$ & $0$ &$\frac{1}{2}$ & $0$ &$0$\\

&$\langle\bar D^{\ast} \Sigma_c^{\ast}, ^4P_{\frac{1}{2}}|\mathscr{L}| \bar D\Sigma_c, ^2P_{\frac{1}{2}}\rangle$ & $0$ &$\frac{1}{2}$ & $0$ &$0$\\

&$\langle\bar D^{\ast} \Sigma_c^{\ast}, ^2P_{\frac{1}{2}}|\mathscr{L}|\bar  D^{\ast}\Sigma_c, ^2P_{\frac{1}{2}}\rangle$ & $-\frac{4\sqrt 2}{81}$ &$\frac{1}{2}$ & $-\frac{8\sqrt 2}{135}$ &$-0.11$\\
&$\langle \bar D^{\ast} \Sigma_c^{\ast}, ^2P_{\frac{1}{2}}|\mathscr{L}| \bar D^{\ast}\Sigma_c, ^4P_{\frac{1}{2}}\rangle$ & $-\frac{4}{81}$ &$\frac{1}{2}$ & $-\frac{8}{135}$ &$-0.08$\\

&$\langle\bar D^{\ast} \Sigma_c^{\ast}, ^4P_{\frac{1}{2}}|\mathscr{L}| \bar D^{\ast}\Sigma_c, ^4P_{\frac{1}{2}}\rangle$ & $\frac{4 \sqrt 5}{81}$&$\frac{1}{2}$ & $\frac{8}{27\sqrt 5}$ &$0.17$\\
&$\langle\bar D^{\ast} \Sigma_c^{\ast}, ^4P_{\frac{1}{2}}|\mathscr{L}|\bar  D\Sigma_c^{\ast}, ^4P_{\frac{1}{2}}\rangle$ & $\frac{1}{18} \sqrt {\frac{5}{3}}$ &$\frac{1}{2}$ & $\frac{1}{3\sqrt{15}}$ &$0.11$\\
&$\langle \bar D^{\ast} \Sigma_c^{\ast}, ^2P_{\frac{1}{2}}|\mathscr{L}|\bar  D\Sigma_c^{\ast}, ^4P_{\frac{1}{2}}\rangle$ & $\frac{1}{9\sqrt 3}$ &$\frac{1}{2}$ & $\frac{2}{{15}\sqrt 3}$ &$0.10$\\
&$\langle \bar D^{\ast} \Sigma_c, ^4P_{\frac{1}{2}}|\mathscr{L}|\bar  D\Sigma_c^{\ast}, ^4P_{\frac{1}{2}}\rangle$ & $0$ &$\frac{1}{2}$ & $0$ &$0$\\
&$\langle \bar D^{\ast} \Sigma_c, ^2P_{\frac{1}{2}}|\mathscr{L}|\bar  D\Sigma_c^{\ast}, ^4P_{\frac{1}{2}}\rangle$ & $0$ &$\frac{1}{2}$ & $0$ &$0$\\

\hline
&&$\langle P^{\frac{3}{2},\frac{3}{2}}_c \pi_0 |\frac{1}{i}\mathscr{L}_{quark}| P'^{\frac{3}{2},\frac{3}{2}}_c  \rangle$/$g_{1}$  &$\langle P^{\frac{3}{2},\frac{3}{2}}_c \pi_0|\frac{1}{i}\mathscr{L}_{eff}| P'^{\frac{3}{2},\frac{3}{2}}_c  \rangle$/$g_{33}$ & $g_{33}$/$g_A$&  $g_{33}$\\
\hline
$\frac{3}{2}^{-}$ & $\langle\bar D\Sigma_c^{\ast}, ^4S_{\frac{3}{2}}|\mathscr{L}|\bar D^{\ast}\Sigma_c, ^4S_{\frac{3}{2}} \rangle$  &$0$ &$\frac{3}{2}$ &0 &0\\
& $\langle\bar D^{\ast}\Sigma_c^{\ast},^4S_{\frac{3}{2}}|\mathscr{L}|\bar D \Sigma_c^{\ast} ,^4S_{\frac{3}{2}} \rangle$ &$\frac{1}{2\sqrt{15}}$  &$\frac{3}{2}$ &$\frac{1}{5\sqrt {15}}$ &0.07\\
& $\langle\bar D^{\ast}\Sigma_c^{\ast},^4S_{\frac{3}{2}} |\mathscr{L}|\bar D^{\ast} \Sigma_c,^4S_{\frac{3}{2}} \rangle$  &$\frac{4}{9\sqrt{5}}$  &$\frac{3}{2}$ & $\frac{8}{45\sqrt5}$ &0.10\\
\hline
&&$\langle P^{\frac{3}{2},\frac{3}{2}}_c \pi_0 |\frac{1}{i}\mathscr{L}_{quark}| P'^{\frac{3}{2},\frac{3}{2}}_c \rangle$/$g_{1}$  &$\langle P^{\frac{3}{2},\frac{3}{2}}_c \pi_0|\frac{1}{i}\mathscr{L}_{eff}| P'^{\frac{3}{2},\frac{3}{2}}_c \pi_0 \rangle$/$g'_{33}$ & $g'_{33}$/$g_A$&  $g'_{33}$\\
\hline
$\frac{3}{2}^{+}$ &$\langle\bar D^{\ast} \Sigma_c, ^2P_{\frac{3}{2}}|\mathscr{L}|\bar  D^{\ast}\Sigma_c, ^4P_{\frac{3}{2}}\rangle$ &$-\frac{19}{27\sqrt 5}$ &$\frac{3}{2}$  &$-\frac{38}{135 \sqrt 5}$ &-0.16\\
 &$\langle\bar D^{\ast} \Sigma_c^{\ast}, ^2P_{\frac{3}{2}}|\mathscr{L}| \bar D^{\ast}\Sigma_c^{\ast}, ^4P_{\frac{3}{2}}\rangle$ &$-\frac{11}{27\sqrt 2}$ &$\frac{3}{2}$  &$-\frac{11\sqrt 2}{135}$ &-0.15\\
&$\langle \bar D^{\ast} \Sigma_c^{\ast}, ^4P_{\frac{3}{2}}|\mathscr{L}| \bar D^{\ast}\Sigma_c^{\ast}, ^6P_{\frac{3}{2}}\rangle$ &$-\frac{11}{25\sqrt 6}$ &$\frac{3}{2}$ &$-\frac{11}{125}\sqrt{\frac{2}{3}}$ &-0.09\\
&$\langle\bar D^{\ast} \Sigma_c^{\ast}, ^6P_{\frac{3}{2}}|\mathscr{L}|\bar  D^{\ast}\Sigma_c^{\ast}, ^2P_{\frac{3}{2}}\rangle$ &$0 $ &$\frac{3}{2}$ &0 &0 \\
&$\langle\bar D^{\ast} \Sigma_c, ^2P_{\frac{3}{2}}|\mathscr{L}|\bar  D\Sigma_c^{\ast}, ^4P_{\frac{3}{2}}\rangle$ &$0$&$\frac{3}{2}$&0 &0\\
&$\langle\bar D^{\ast} \Sigma_c, ^4P_{\frac{3}{2}}|\mathscr{L}|\bar  D\Sigma_c^{\ast}, ^4P_{\frac{3}{2}}\rangle$ &$0$ &$\frac{3}{2}$ &0 &0\\
&$\langle\bar D^{\ast} \Sigma_c^{\ast}, ^6P_{\frac{3}{2}}|\mathscr{L}|  \bar D\Sigma_c^{\ast}, ^4P_{\frac{3}{2}}\rangle$ &$-\frac{1}{5 \sqrt{10}}$ &$\frac{3}{2}$ &$-\frac{1}{25}\sqrt{\frac{2}{5}}$ &-0.03\\

&$\langle\bar D^{\ast} \Sigma_c^{\ast}, ^4P_{\frac{3}{2}}|\mathscr{L}|\bar  D^{\ast}\Sigma_c, ^4P_{\frac{3}{2}}\rangle$ & $\frac{44}{135\sqrt 5}$ &$\frac{3}{2}$  & $\frac{88}{675\sqrt5}$ &0.07\\
& $\langle\bar D^{\ast}\Sigma_c^{\ast},^4P_{\frac{3}{2}}|\mathscr{L}|\bar D \Sigma_c^{\ast} ,^4P_{\frac{3}{2}} \rangle$ &$\frac{11}{30\sqrt{15}}$  &$\frac{3}{2}$ &$\frac{11}{75\sqrt {15}}$ &0.05\\

\hline
&&$\langle P^{\frac{5}{2},\frac{5}{2}}_c \pi_0|\frac{1}{i}\mathscr{L}_{quark}| P'^{\frac{5}{2},\frac{5}{2}}_c \rangle$/$g_{1}$  &$\langle P^{\frac{5}{2},\frac{5}{2}}_c \pi_0 |\frac{1}{i}\mathscr{L}_{eff}| P'^{\frac{5}{2},\frac{5}{2}}_c \pi_0 \rangle$/$g_{55}$ & $g_{55}$ & $g_{55}$\\
\hline
$\frac{5}{2}^{+}$ &$\langle\bar D^{\ast} \Sigma_c^{\ast}, ^4P_{\frac{5}{2}}|\mathscr{L}|  \bar D^{\ast} \Sigma_c^{\ast}, ^6P_{\frac{5}{2}}\rangle$ &$-\frac{11}{15\sqrt{21}}$ &$\frac{5}{2}$ &$-\frac{22}{125\sqrt{21}}$ &-0.05\\
&$\langle\bar D^{\ast} \Sigma_c, ^4P_{\frac{5}{2}}|\mathscr{L}| \bar D^{\ast} \Sigma_c^{\ast}, ^6P_{\frac{5}{2}}\rangle$ &$\frac{4}{3\sqrt{105}}$  &$\frac{5}{2}$  &$\frac{8}{25\sqrt{105}}$ &0.04\\
&$\langle\bar D^{\ast} \Sigma_c, ^4P_{\frac{5}{2}}|\mathscr{L}| \bar D^{\ast} \Sigma_c^{\ast}, ^4P_{\frac{5}{2}}\rangle$ &$\frac{4}{9\sqrt{5}}$  &$\frac{5}{2}$  &$\frac{8}{75\sqrt 5}$ &0.06\\
&$\langle\bar D \Sigma_c^{\ast}, ^4P_{\frac{5}{2}}|\mathscr{L}| \bar D^{\ast} \Sigma_c^{\ast}, ^6P_{\frac{5}{2}}\rangle$ &$-\frac{1}{3\sqrt{35}}$  &$\frac{5}{2}$   &$-\frac{2}{25\sqrt{35}}$  &-0.02\\
&$\langle\bar D \Sigma_c^{\ast}, ^4P_{\frac{5}{2}}|\mathscr{L}|\bar  D^{\ast} \Sigma_c^{\ast}, ^4P_{\frac{5}{2}}\rangle$ &$\frac{1}{2\sqrt{15}}$ &$\frac{5}{2}$   &$\frac{\sqrt 3}{25\sqrt {5}}$ &0.04\\
&$\langle\bar D \Sigma_c^{\ast}, ^4P_{\frac{5}{2}}|\mathscr{L}|\bar  D^{\ast} \Sigma_c, ^4P_{\frac{5}{2}}\rangle$ &$0$ &$\frac{5}{2}$ &$0$ &0\\
\bottomrule[1pt]\bottomrule[1pt]
      \end{tabular}
  \end{center}
\end{table*}

The Lagrangians for $P'_c(J^P=\frac{3}{2}^{\pm})\rightarrow
P_c(J^P=\frac{1}{2}^{\pm}) \pi_0$ and
$P'_c(J^P=\frac{5}{2}^{\pm})\rightarrow P_c(J^P=\frac{3}{2}^{\pm})
\pi_0$ are
 \begin{eqnarray} \label{w17}
&&\mathscr{L}^{\frac{3}{2}\rightarrow\frac{1}{2}}_{eff}=\frac{1}{2}g_{23} \bar P_c \partial_{\mu}\phi P'^{\mu}_c \sim \frac{1}{2}\frac{g_{{23}}}{f_\pi}\bar P_{c} \partial^{\mu}\pi_0P' _{c\mu}\nonumber \\
&&\mathscr{L}^{\frac{5}{2}\rightarrow
\frac{3}{2}}_{eff}=\frac{1}{2}g_{{35}} \bar P^{\nu}_{c }
\partial^{\mu} \phi P'_{c \mu \nu} \sim \frac{1}{2} \frac{g_{{35}}}{f_\pi}  \bar
P_{c}^\nu \partial^{\mu}\pi_0 P'_{c \mu \nu},
\end{eqnarray}
where $g_{23}$ and $g_{35}$ denote the $P'_c(\frac{3}{2}^{-}) P_c(\frac{1}{2}^{-}) \pi_0$ and $P'_c(\frac{5}{2}^{-})P_c(\frac{3}{2}^{-}) \pi_0$ coupling constants respectively. We use the $g'_{23}$ and $g'_{35}$ to represent the couplings when the two pentaquark states have positive parities. %with the positive parity pentaquark states.
When these pentaquark states are in the $2_{2f}$ representation with
$(I,I_3)=(\frac{1}{2},\frac{1}{2})$, we have
\begin{eqnarray} \label{w18}
&&\langle P^{\frac{1}{2},\frac{1}{2}}_c \pi_0   |\frac1i\mathscr{L}^{\frac{3}{2}\rightarrow\frac{1}{2}}_{eff}|P'^{\frac{3}{2},\frac{1}{2}}_c \rangle =\frac{1}{\sqrt 6} \frac{q_z}{f_\pi} g_{23}, \nonumber\\
&&\langle P^{\frac{3}{2},\frac{3}{2}}_c \pi_0 |\frac1i \mathscr{L}^{\frac{5}{2}\rightarrow
\frac{3}{2}}_{eff}|P'^{\frac{5}{2},\frac{3}{2}}_c  \rangle
=\frac{1}{2}\sqrt{\frac{2}{5}}\frac{q_z}{f_\pi} g_{35}.
\end{eqnarray}

We collect the numerical results in Table \ref{rdc}. Several
coupling constants vanish due to the same mechanism as in Table
\ref{rdcs}. When the pentaquark states have different $J$, the
numerical results for $P'_cP_c\pi_0$ are lager than those when they
have the same $J^P$ but different configurations.

\renewcommand{\arraystretch}{1.5}
\begin{table*}[htbp]
\caption{The transition coupling constants of the molecular
pentaquark states at the quark level and the hadron level. The two
pentaquark states are in the $2_{2f}$ representation with
$(I,I_3)=(\frac{1}{2},\frac{1}{2})$. The $P'_c$ and $P_c$ have
different $J^P$ quantum numbers. The superscripts denote the total
angular momentum and its third component. }\label{rdc}
\begin{center}
   \begin{tabular}{c|c|c|c|c|c} \toprule[1pt]\toprule[1pt]
&&$\langle P^{\frac{3}{2},\frac{1}{2}}_c \pi_0 |\frac{1}{i}\mathscr{L}_{quark}| P'^{\frac{1}{2},\frac{1}{2}}_c  \rangle$/$g_{1}$  &$\langle P^{\frac{3}{2},\frac{1}{2}}_c \pi_0 |\frac{1}{i}\mathscr{L}_{eff}| P'^{\frac{1}{2},\frac{1}{2}}_c  \rangle$/$g_{23}$ & $g_{23}$/$g_A$&  $g_{23}$\\
\hline
$\frac{1}{2}^{-}\rightarrow \frac{3}{2}^{-}$ & $\langle \bar D\Sigma_c^{\ast}, ^4S_{\frac{3}{2}}|\mathscr{L}|\bar D \Sigma_c, ^2S_{\frac{1}{2}} \rangle$  &$-\frac{2\sqrt2}{9}$ &$\frac{1}{\sqrt 6}$ &-$\frac{4}{5\sqrt 3}$ &-0.59\\
& $\langle \bar D^{\ast}\Sigma_c, ^4S_{\frac{3}{2}}|\mathscr{L}|\bar D\Sigma_c, ^2S_{\frac{1}{2}} \rangle$  &$\frac{1}{3\sqrt 6}$ &$\frac{1}{\sqrt 6}$ &$-{\frac{1}{5}}$ &-0.25 \\

& $\langle \bar  D^{\ast}\Sigma_c^{\ast}, ^4S_{\frac{3}{2}}|\mathscr{L}| \bar D\Sigma_c , ^2S_{\frac{1}{2}} \rangle$  &$0$ &$\frac{1}{\sqrt 6}$ &$0$ &0\\
\hline
&&$\langle P^{\frac{3}{2},\frac{1}{2}}_c \pi_0|\frac{1}{i}\mathscr{L}_{quark}| P'^{\frac{1}{2},\frac{1}{2}}_c \rangle$/$g_{1}$  &$\langle P^{\frac{3}{2},\frac{1}{2}}_c \pi_0|\frac{1}{i}\mathscr{L}_{eff}| P'^{\frac{1}{2},\frac{1}{2}}_c  \rangle$/$g'_{23}$ & $g'_{23}$/$g_A$&  $g'_{23}$\\
\hline
$\frac{1}{2}^{+}\rightarrow \frac{3}{2}^{+}$ & $\langle \bar D^{\ast}\Sigma_c, ^2P_{\frac{3}{2}}|\mathscr{L}|\bar D \Sigma_c, ^2P_{\frac{1}{2}} \rangle$  &$\frac{1}{9}\sqrt{\frac{2}{3}}$ &$\frac{1}{\sqrt 6}$   &$\frac{2}{15}$ &0.17\\
 & $\langle\bar D^{\ast}\Sigma_c, ^4P_{\frac{3}{2}}|\mathscr{L}|\bar D \Sigma_c, ^2P_{\frac{1}{2}} \rangle$  &$\frac{1}{9}\sqrt{\frac{5}{6}}$ &$\frac{1}{\sqrt 6}$   &$\frac{1}{3\sqrt  5}$ &0.19\\
 & $\langle\bar D^{\ast}\Sigma_c, ^2P_{\frac{3}{2}}|\mathscr{L}|\bar D^{\ast} \Sigma_c, ^2P_{\frac{1}{2}} \rangle$  &$-\frac{14\sqrt 2}{81}$ &$\frac{1}{\sqrt 6}$   &$-\frac{28}{45\sqrt 3}$ &-0.46\\
 & $\langle\bar D^{\ast}\Sigma_c, ^4P_{\frac{3}{2}}|\mathscr{L}|\bar D^{\ast} \Sigma_c, ^2P_{\frac{1}{2}} \rangle$  &$\frac{19\sqrt 5}{81}$ &$\frac{1}{\sqrt 6}$   &$\frac{19}{9\sqrt {15}}$ &0.69\\
 & $\langle\bar D^{\ast}\Sigma_c, ^4P_{\frac{3}{2}}|\mathscr{L}|\bar D^{\ast} \Sigma_c, ^4P_{\frac{1}{2}} \rangle$  &$\frac{5\sqrt 5}{81}$ &$\frac{1}{\sqrt 6}$   &$\frac{1}{9}\sqrt{\frac{10}{3}}$ &0.26\\
 & $\langle\bar D\Sigma_c^{\ast}, ^4P_{\frac{3}{2}}|\mathscr{L}|\bar D \Sigma_c^{\ast}, ^4P_{\frac{1}{2}} \rangle$  &$\frac{4\sqrt 5}{27}$ &$\frac{1}{\sqrt 6}$   &$\frac{4}{3}\sqrt{\frac{2}{15}}$ &0.62\\
 & $\langle\bar D\Sigma_c^{\ast}, ^4P_{\frac{3}{2}}|\mathscr{L}|\bar D^{\ast }\Sigma_c, ^4P_{\frac{1}{2}} \rangle$  &$0$ &$\frac{1}{\sqrt 6}$   &$0$ &0\\
 & $\langle\bar D\Sigma_c^{\ast}, ^4P_{\frac{3}{2}}|\mathscr{L}|\bar D^{\ast} \Sigma_c, ^2P_{\frac{1}{2}} \rangle$  &$0$ &$\frac{1}{\sqrt 6}$   &$0$ &0\\

 & $\langle\bar D^{\ast}\Sigma_c^{\ast}, ^2P_{\frac{3}{2}}|\mathscr{L}|\bar D^{\ast }\Sigma_c, ^2P_{\frac{1}{2}} \rangle$  &$\frac{16}{81}$ &$\frac{1}{\sqrt 6}$   &$\frac{16}{45}\sqrt {\frac{2}{3}}$ &0.37\\
 & $\langle\bar D^{\ast}\Sigma_c^{\ast}, ^2P_{\frac{3}{2}}|\mathscr{L}|\bar D^{\ast} \Sigma_c, ^4P_{\frac{1}{2}} \rangle$  &$-\frac{\sqrt 2}{81}$ &$\frac{1}{\sqrt 6}$   &$-\frac{2}{45\sqrt 3}$ &-0.03\\

& $\langle\bar D^{\ast}\Sigma_c^{\ast}, ^2P_{\frac{3}{2}}|\mathscr{L}|\bar D^{\ast}\Sigma_c^{\ast}, ^2P_{\frac{1}{2}} \rangle$  &$\frac{23\sqrt 2}{81}$ &$\frac{1}{\sqrt 6}$ &$\frac{46}{45\sqrt{3}}$  &0.75\\
& $\langle\bar D^{\ast}\Sigma_c^{\ast}, ^4P_{\frac{3}{2}}|\mathscr{L}|\bar D^{\ast}\Sigma_c^{\ast}, ^2P_{\frac{1}{2}} \rangle$  &$\frac{55}{162}$ &$\frac{1}{\sqrt 6}$  &$\frac{11}{9\sqrt 6}$  &0.63\\
& $\langle\bar D^{\ast}\Sigma_c^{\ast}, ^4P_{\frac{3}{2}}|\mathscr{L}|\bar D^{\ast}\Sigma_c^{\ast}, ^4P_{\frac{1}{2}} \rangle$  &$\frac{38}{81\sqrt5}$ &$\frac{1}{\sqrt 6}$  &$\frac{38}{45}\sqrt {\frac{2}{15}}$  &0.39\\
\hline
&&$\langle P^{\frac{3}{2},\frac{3}{2}}_c\pi_0  |\frac{1}{i}\mathscr{L}_{quark}| P'^{\frac{5}{2},\frac{3}{2}}_c \rangle$/$g_1$  &$\langle P^{\frac{3}{2},\frac{3}{2}}_c \pi_0|\frac{1}{i}\mathscr{L}_{eff}|  P'^{\frac{5}{2},\frac{3}{2}}_c\rangle$/$g_{35}$ & $g_{35}$/$g_A$&  $g'_{35}$\\
\hline

$\frac{5}{2}^{-}\rightarrow \frac{3}{2}^{-}$ & $\langle\bar D\Sigma_c^{\ast}, ^4S_{\frac{3}{2}}|\mathscr{L}| \bar D^{\ast}\Sigma_c^{\ast}, ^6S_{\frac{5}{2}} \rangle$  &$\frac{1}{3\sqrt{10}}$ &$\frac{1}{2}\sqrt {\frac{2}{5}}$ &$\frac{1}{5}$ &0.25\\
& $\langle\bar D^{\ast}\Sigma_c, ^4S_{\frac{3}{2}}|\mathscr{L}| \bar D^{\ast}\Sigma_c^{\ast}, ^6S_{\frac{5}{2}} \rangle$  &$-\frac{2}{3}\sqrt{\frac{2}{15}}$ &$\frac{1}{2}\sqrt {\frac{2}{5}}$ &$-\frac{4}{5\sqrt3}$ &-0.59\\

& $\langle \bar D^{\ast}\Sigma_c^{\ast}, ^4S_{\frac{3}{2}}|\mathscr{L}|\bar D^{\ast}\Sigma_c^{\ast}, ^6S_{\frac{5}{2}} \rangle$  &$\frac{11}{15\sqrt 6}$&$\frac{1}{2}\sqrt {\frac{2}{5}}$  & $\frac{11}{5\sqrt {15}}$&0.72\\
\hline

&&$\langle P^{\frac{3}{2},\frac{3}{2}}_c \pi_0  |\frac{1}{i}\mathscr{L}_{quark}| P'^{\frac{5}{2},\frac{3}{2}}_c \rangle$/$g_1$  &$\langle  P^{\frac{3}{2},\frac{3}{2}}_c \pi_0 |\frac{1}{i}\mathscr{L}_{eff}|P'^{\frac{5}{2},\frac{3}{2}}_c \rangle$/$g'_{35}$ & $g'_{35}$/$g_A$&  $g'_{35}$\\
\hline

$ \frac{3}{2}^{+} \rightarrow \frac{5}{2}^{+}$ &$\langle \bar D^{\ast} \Sigma_c, ^4P_{\frac{5}{2}}|\mathscr{L}|\bar  D\Sigma_c, ^2P_{\frac{3}{2}}\rangle$  &$\frac{1}{3\sqrt{10}}$ &$\frac{1}{2}\sqrt {\frac{2}{5}}$ &$\frac{1}{5}$ &0.25\\
&$\langle\bar D^{\ast} \Sigma_c, ^4P_{\frac{5}{2}}|\mathscr{L}|\bar  D\Sigma_c^{\ast}, ^4P_{\frac{3}{2}}\rangle$ &$0$ &$\frac{1}{2}\sqrt {\frac{2}{5}}$ &0 &0\\

&$\langle\bar D^{\ast} \Sigma_c^{\ast}, ^2P_{\frac{3}{2}}|\mathscr{L}|\bar  D^{\ast}\Sigma_c^{\ast}, ^6P_{\frac{5}{2}}\rangle$ &0 &$\frac{1}{2}\sqrt {\frac{2}{5}}$  & 0 &0 \\

&$\langle\bar D^{\ast} \Sigma_c^{\ast}, ^2P_{\frac{3}{2}}|\mathscr{L}|\bar  D^{\ast}\Sigma_c^{\ast}, ^4P_{\frac{5}{2}}\rangle$ &$\frac{11}{18\sqrt 3}$ &$\frac{1}{2}\sqrt {\frac{2}{5}}$ &$\frac{11}{3\sqrt{30}}$   &0.85\\

&$\langle\bar D^{\ast} \Sigma_c^{\ast}, ^4P_{\frac{3}{2}}|\mathscr{L}| \bar D^{\ast}\Sigma_c^{\ast}, ^6P_{\frac{5}{2}}\rangle$ &$\frac{11}{75}\sqrt{\frac{7}{2}}$ &$\frac{1}{2}\sqrt {\frac{2}{5}}$  &$\frac{11}{25}\sqrt{\frac{7}{5}}$ &0.66\\
&$\langle\bar D^{\ast} \Sigma_c^{\ast}, ^4P_{\frac{3}{2}}|\mathscr{L}|\bar
D^{\ast}\Sigma_c^{\ast}, ^4P_{\frac{5}{2}}\rangle$
&$\frac{38}{225}\sqrt{\frac{2}{3}}$ &$\frac{1}{2}\sqrt
{\frac{2}{5}}$

&$\frac{75}{76\sqrt {15}}$ &0.33\\
&$\langle\bar D^{\ast} \Sigma_c^{\ast}, ^6P_{\frac{3}{2}}|\mathscr{L}| \bar D^{\ast}\Sigma_c^{\ast}, ^6P_{\frac{5}{2}}\rangle$ &$\frac{2}{25} \sqrt{\frac{7}{3}}$ &$\frac{1}{2}\sqrt {\frac{2}{5}}$ &$\frac{2}{25}\sqrt {\frac{42}{5}}$ &0.29 \\

%\hline$ \frac{5}{2}^{+} \rightarrow \frac{3}{2}^{+}$
&$\langle \bar D^{\ast} \Sigma_c, ^2P_{\frac{3}{2}}|\mathscr{L}| \bar D^{\ast}\Sigma_c^{\ast}, ^6P_{\frac{5}{2}}\rangle$ &$0$&$\frac{1}{2}\sqrt {\frac{2}{5}}$  & 0 &0 \\
&$\langle\bar D^{\ast} \Sigma_c, ^2P_{\frac{3}{2}}|\mathscr{L}| \bar D^{\ast}\Sigma_c^{\ast}, ^4P_{\frac{5}{2}}\rangle$ &$-\frac{2}{9}\sqrt{\frac{2}{3}}$ &$\frac{1}{2}\sqrt {\frac{2}{5}}$ & $-\frac{4}{3\sqrt {15}}$ &-0.44\\
&$\langle\bar D^{\ast} \Sigma_c, ^4P_{\frac{3}{2}}|\mathscr{L}| \bar D^{\ast}\Sigma_c^{\ast}, ^6P_{\frac{5}{2}}\rangle$ &$-\frac{2}{15}\sqrt{\frac{14}{5}}$&$\frac{1}{2}\sqrt {\frac{2}{5}}$  &$-\frac{4\sqrt 7}{25}$ &-0.54\\

&$\langle\bar D^{\ast} \Sigma_c, ^4P_{\frac{3}{2}}|\mathscr{L}| \bar D^{\ast}\Sigma_c^{\ast}, ^4P_{\frac{5}{2}}\rangle$ &$\frac{8}{45}\sqrt{\frac{2}{15}}$ &$\frac{1}{2}\sqrt {\frac{2}{5}}$  &$\frac{16}{75\sqrt 3}$  &0.16\\

&$\langle\bar D^{\ast} \Sigma_c, ^4P_{\frac{3}{2}}|\mathscr{L}| \bar D^{\ast}\Sigma_c, ^4P_{\frac{5}{2}}\rangle$ &$\frac{1}{9}\sqrt{\frac{2}{3}}$ &$\frac{1}{2}\sqrt {\frac{2}{5}}$  &$\frac{2}{3\sqrt {15}}$  &0.22\\

&$\langle\bar D^{\ast} \Sigma_c, ^2P_{\frac{3}{2}}|\mathscr{L}| \bar D^{\ast}\Sigma_c, ^4P_{\frac{5}{2}}\rangle$ &$\frac{19}{9\sqrt {30}}$ &$\frac{1}{2}\sqrt {\frac{2}{5}}$  &$\frac{19}{15\sqrt 3}$  &0.93\\
\bottomrule[1pt]\bottomrule[1pt]
  \end{tabular}
   \end{center}
  \end{table*}

%%%%%%%%%%%%%%%%%%%%%%%%%%%%%%%%%%%%
\section{The suppression of the axial charges of
some hidden-charm pentaquark states}\label{sec5}
%%%%%%%%%%%%%%%%%%%%%%%%%%%%%%%%%%%%

We notice that the axial charges of the hidden-charm pentaquarks are
generally smaller than the axial charge of the proton. For
comparison, let us take the proton as an example. The color wave
function of the three quarks within the proton are antisymmetric.
The spin and flavor wave functions of the three light quarks are
totally symmetric, which may ensure the contribution to the axial
charge from three quarks and different components in the spin and
flavor wave functions are constructive. In contrast, the spin and
the flavor wave functions of $q_1$ and $q_2q_3$ are not totally
symmetric. The interference between the contributions from different
components may lead to a small axial charge. In the following, we
pick out the component in the spin and flavor functions of the $P_c$
states where the three quarks form a color singlet $s$-wave octet or
decuplet baryon. Then we calculate their axial charges and compare
with those of the usual $P_c$ states.

We first decompose the molecular $P_c$ states as follows:
\begin{eqnarray} \label{w19}
&&\Big|\Big[( c^a (q^b_2 q^f_3)_{s_{23}})_{s_t}\otimes {(\bar c^d q^e_1)_{s_{1Q}}} \Big]_{s}\otimes l \Big\rangle^{J_z} _J \delta_{de}\epsilon_{abf}\nonumber\\
&&=\sum_{s'_d,s'_t}  {\hat{s}}_t {\hat{s}_{1Q}} {\hat{s}}'_d
{\hat{s}}'_t \left(
  \begin{array}{ccc}
    \frac{1}{2}_{c} &s_{23}  &s_t\\
     \frac{1}{2}_{\bar c} &\frac{1}{2}_{q_1} &s_{1Q}\\
     s'_d  &s'_t &s\\
 \end{array}
\right)\Big|\Big[( \frac{1}{2}_{c^a}\otimes \frac{1}{2}_{\bar
c^d})_{s'_d}\otimes  \{ (q^b_2 q^f_3)_{s_{23}}\otimes
q^e_1\}_{s'_t} \Big]_{s}\otimes l \Big\rangle^{J_z} _J\frac{1}{\sqrt
3}(\delta_{da}\epsilon_{ebf}+\delta_{db}\epsilon_{aef}+\delta_{df}\epsilon_{abe}),
\end{eqnarray}
where $a$, $b$, $d$, $e$ and $f$ are the color index, $s'_d$ is the spin of
the charmonium, $s'_t$ is the total spin of the three light quarks.
${\hat{s}}_t=\sqrt{2s_t+1}$. If the $q_1q_2q_3$ form a color
singlet s-wave baryon, their isospin and spin are either
$\frac{3}{2}$ or $\frac{1}{2}$.

Let us take the $\frac{3}{2}^-$ and $\frac{5}{2}^-$ molecular states
%$|^6S_{\frac{5}{2}} ,\bar D^{\ast}
%\Sigma_c^{\ast}\rangle$ and $|^4S_{\frac{3}{2}},
%\bar D\Sigma_c^{\ast}\rangle$
with the isospin
$(I,I_3)=(\frac{1}{2},\frac{1}{2})$ as an example:
\begin{eqnarray} \label{w20}
&&|^6S_{\frac{5}{2}}, \bar D^{\ast} \Sigma_c^{\ast}\rangle=\Big|\Big[(
\frac{1}{2}_{c^a}\otimes \frac{1}{2}_{\bar c^d})_{1}\otimes  \{
(q^b_2 q^f_3)_{1}\otimes q^e_1\}_{\frac{3}{2}}
\Big]_{\frac{5}{2}}\otimes 0 \Big\rangle^{J_z} _J\frac{1}{\sqrt
3}(\delta_{da}\epsilon_{ebf}+\delta_{db}\epsilon_{aef}+\delta_{df}\epsilon_{abe}),
\end{eqnarray}
The three quarks $q_1q_2q_3$ in this $\frac{5}{2}^-$ pentaquark
state can not form an $s$-wave baryon because their total spin is
$\frac{3}{2}$ and their total isospin is $\frac{1}{2}$. Its axial charge is $0.15$.

For the  $|^4S_{\frac{3}{2}}, D\Sigma_c^{\ast}\rangle$ state,
\begin{eqnarray} \label{w21}
&&|^4S_{\frac{3}{2}}, \bar D\Sigma_c^{\ast}\rangle =\Big(\frac{1}{\sqrt
3}\Big|\Big[( \frac{1}{2}_{c^a}\otimes \frac{1}{2}_{\bar
c^d})_{1}\otimes \{ (q^b_2\otimes q^f_3)_{1}\otimes
q^e_1\}_{\frac{1}{2}}\Big]_{\frac{3}{2}}\otimes l \Big\rangle-\frac{1}{2}\Big|\Big[(
\frac{1}{2}_{c^a}\otimes \frac{1}{2}_{\bar c^d})_{0}\otimes \{
(q^b_2\otimes q^f_3)_{1}\otimes
q^e_1\}_{\frac{3}{2}}\Big]_{\frac{3}{2}}\otimes l \Big\rangle \nonumber\\
&&+\frac{\sqrt 5}{2\sqrt 3}\Big|\Big[( \frac{1}{2}_{c^a}\otimes
\frac{1}{2}_{\bar c^d})_{1}\otimes  \{(q^b_2\otimes
q^f_3)_{1}\otimes q^e_1\}_{\frac{3}{2}} \Big]_{\frac{3}{2}}\otimes l
\Big\rangle\Big ) \frac{1}{\sqrt
3}(\delta_{da}\epsilon_{ebf}+\delta_{db}\epsilon_{afe}+\delta_{df}\epsilon_{abe})
\end{eqnarray}
The component in which the three light quarks $q_1q_2q_3$ form an
$s$-wave baryon reads
\begin{eqnarray} \label{w220}
&&P_{cs}=\Big|\Big[( \frac{1}{2}_{c^a}\otimes \frac{1}{2}_{\bar
c^d})_{1}\otimes \{ (q^b_2\otimes q^f_3)_{1}\otimes
q^e_1\}_{\frac{1}{2}} \Big]_{\frac{3}{2}}\otimes l
\Big\rangle\delta_{da}\epsilon_{ebf}.
\end{eqnarray}
The spin and flavor wave function of $P_{cs}$ is
\begin{eqnarray} \label{w22}
&&P_{cs}=\frac{\sqrt 2}{3}(u\uparrow u\uparrow d\downarrow+u\uparrow d\downarrow u\uparrow+d\downarrow u\uparrow u\uparrow\nonumber\\
&&-\frac{1}{2}u\downarrow d\uparrow u\uparrow-\frac{1}{2}d\uparrow
u\downarrow u\uparrow-\frac{1}{2}u\uparrow u\downarrow
d\uparrow-\frac{1}{2}u\downarrow u\uparrow
d\uparrow-\frac{1}{2}u\uparrow d\uparrow
u\downarrow-\frac{1}{2}d\uparrow u\uparrow
u\downarrow)\cdot(c\uparrow \bar c \uparrow).
\end{eqnarray}

The spin-flavor wave function of the light quarks
in $P_{cs}$ are the same as that of the proton. The three
quarks are totally symmetric in the spin-flavor space.
The pionic couplings of the $P_{cs}$ are always the same at the quark level.
The $P_{cs}P_{cs}\pi_0$ coupling is $0.63$, which is clearly not
suppressed. The possibility of the $P_{cs}$ component is
$\frac{1}{9}$ in the $\frac{3}{2}^{-}$ molecular state $|\bar D
\Sigma_c^{\ast}\rangle$. The axial charge of the $\frac{3}{2}^{-}$
state $|\bar D \Sigma_c^{\ast}\rangle$ is $0.34$, while the $P_{c}P_{c}\pi_0$ coupling is $0.51$.

The $\frac{3}{2}^{-}$ pentaquark state contains two other possible
configurations $|\bar D^* \Sigma_c\rangle$ and $|\bar D^*
\Sigma^*_c\rangle$. They can be written as
\begin{eqnarray} \label{w23}
&&|^4S_{\frac{3}{2}}, \bar D^{\ast}\Sigma_c\rangle =\Big(-\frac{1}{3}\Big|\Big[( \frac{1}{2}_{c^a}\otimes \frac{1}{2}_{\bar
c^d})_{1}\otimes \{ (q^b_2\otimes q^f_3)_{1}\otimes
q^e_1\}_{\frac{1}{2}}\Big]_{\frac{3}{2}}\otimes l \Big\rangle+\frac{1}{\sqrt 3}\Big|\Big[(
\frac{1}{2}_{c^a}\otimes \frac{1}{2}_{\bar c^d})_{0}\otimes \{
(q^b_2\otimes q^f_3)_{1}\otimes
q^e_1\}_{\frac{3}{2}}\Big]_{\frac{3}{2}}\otimes l \Big\rangle \nonumber\\
&&+\frac{\sqrt 5}{3}\Big|\Big[( \frac{1}{2}_{c^a}\otimes
\frac{1}{2}_{\bar c^d})_{1}\otimes  \{(q^b_2\otimes
q^f_3)_{1}\otimes q^e_1\}_{\frac{3}{2}} \Big]_{\frac{3}{2}}\otimes l
\Big\rangle\Big )\frac{1}{\sqrt
3}(\delta_{da}\epsilon_{ebf}+\delta_{db}\epsilon_{afe}+\delta_{df}\epsilon_{abe}).
\end{eqnarray}

\begin{eqnarray} \label{w24}
&&|^4S_{\frac{3}{2}}, \bar D^{\ast}\Sigma_c^{\ast}\rangle =\Big(\frac{\sqrt 5}{3}\Big|\Big[( \frac{1}{2}_{c^a}\otimes \frac{1}{2}_{\bar
c^d})_{1}\otimes \{ (q^b_2\otimes q^f_3)_{1}\otimes
q^e_1\}_{\frac{1}{2}}\Big]_{\frac{3}{2}}\otimes l \Big\rangle+\frac{1}{2}\sqrt{\frac{5}{3}}\Big|\Big[(
\frac{1}{2}_{c^a}\otimes \frac{1}{2}_{\bar c^d})_{0}\otimes \{
(q^b_2\otimes q^f_3)_{1}\otimes
q^e_1\}_{\frac{3}{2}}\Big]_{\frac{3}{2}}\otimes l \Big\rangle \nonumber\\
&&-\frac{1}{6}\Big|\Big[( \frac{1}{2}_{c^a}\otimes
\frac{1}{2}_{\bar c^d})_{1}\otimes  \{(q^b_2\otimes
q^f_3)_{1}\otimes q^e_1\}_{\frac{3}{2}} \Big]_{\frac{3}{2}}\otimes l
\Big\rangle\Big ) \frac{1}{\sqrt
3}(\delta_{da}\epsilon_{ebf}+\delta_{db}\epsilon_{afe}+\delta_{df}\epsilon_{abe}).
\end{eqnarray}

The possibilities of the $P_{cs}$ component are $\frac{1}{27}$ and $\frac{5}{27}$
in the two states, respectively. However, their axial charges are $0.14$ and $0.21$,
respectively, and much smaller than that of the proton.
The interference of the contributions
from different components reduce the axial charges of the $P_c$
states.

Note that the decomposition in Eq. (\ref{w22}) holds for the
$s$-wave states only. Once there is a $p$-wave excitation between the
meson and the baryon, the above simple symmetry analysis does not apply
any more.

%%%%%%%%%%%%%%%%%%%%%%%%%%%%%%%%%%%%%%%%
\section{Summary}\label{sec6}
%%%%%%%%%%%%%%%%%%%%%%%%%%%%%%%%%%%%%%%%

The observation of the hidden-charm pentaquarks provides a new
platform to study the exotic states in QCD. The axial charges of the
$P_c$ states are very sensitive to the color, spin, and flavor
configurations of the light quarks. In this work, we have derived
the axial charges of the hidden-charm pentaquarks in different
models systematically in the framework of the chiral quark model.

We first construct the color-spin-flavor wave functions of the $P_c$
states under the SU(2) flavor symmetry. The observed two $P_c$
states with the isospin $(I,I_3)=(\frac{1}{2},\frac{1}{2})$ are in
either the $2_{2f}$ or the $2_{1f}$ representations. At the quark
level, we use the chiral quark model to calculate the pionic
couplings for the $P'_cP_c\pi_0$ interaction in the molecular, the
diquark-triquark, and the diquark-diquark-antiquark models. We derive
the analytical expressions of the axial charges of the $P_c$ states
with various quantum numbers.

We notice that the $P_c$ states with the same $J^P$ quantum numbers
may have very different color-spin-flavor wave functions, which
result in different inner angular momentum configurations of the
light quarks and very different axial charges of the pentaquarks. In
other words, the axial charges of the $P_c$ states encode important
information on their underlying structures.

The axial charges of the hidden-charm pentaquarks are generally
smaller than the axial charge of the proton. Within the proton, the
spin-flavor wave functions of the three light quarks are totally
symmetric while their color wave function is totally antisymmetric.
In contrast, the color wave function of the three light quarks
within the $P_c$ states is not necessarily antisymmetric. There
exists interference between the contributions to the $P_c$ axial
charges from the three light quarks even for the $s$-wave $P_c$ states
without orbital excitations. Sometimes such contributions are even
destructive, which renders a small axial charge of the $P_c$ states.
Hopefully, the axial charges of the hidden-charm pentaquarks may be
measured through the semileptonic decays in the near future.

%%%%%%%%%%%%%%%%%%%%%%%%%%%%%%%%%%%%%%%
\section*{Acknowledgments}
%%%%%%%%%%%%%%%%%%%%%%%%%%%%%%%%%%%%%%%

This project is supported by the National Natural Science Foundation
of China under Grants No. 11261130311, No. 11575008 and 973 program.

%%%%%%%%%%%%%%%%%%%%%%%%%%%%%%%
\section*{Appendix}\label{appendix}
%%%%%%%%%%%%%%%%%%%%%%%%%%%%%%%

The recently observed two $P_c$ states have the isospin
$(I,I_3)=({\frac{1}{2},\frac{1}{2}})$. In the SU(3) flavor
symmetry limit, these pentaquark states belong to the $8_f$ flavor
representation. More details can be seen in Ref.
\cite{Wang:2016dzu}. The SU(3) invariant Lagrangian at the hadron
level reads
\begin{eqnarray} \label{a1}
\mathscr{L}^{8_f}_{eff}={\rm Tr}\Big(\bar{P_c}
\gamma^{\mu}\gamma^5(g\{\partial ^{\mu}\phi,P_c\}+f[\partial
^{\mu}\phi,P_c])\Big),
\end{eqnarray}
where $g$ and $f$ are two independent couplings, and %$u^{\mu}$ is the same as Eq. (\ref{w5}).
\begin{equation} \label{a2}
P_c=\left(
  \begin{array}{ccc}
    \frac{P_{(1,0,-1)}}{\sqrt 2}+\frac{P{(0,0,-1)}}{\sqrt 6} &P_{(1,1,-1)}  &P_{(\frac{1}{2},\frac{1}{2},0)}\\
     P_{(1,-1,-1)} &-\frac{P_{(1,0,-1)}}{\sqrt 2}+\frac{P_{(0,0,-1)}}{\sqrt 6}  &P_{(\frac{1}{2},-\frac{1}{2},0)}\\
     P_{(\frac{1}{2},-\frac{1}{2},-2)} &P_{(\frac{1}{2},\frac{1}{2},-2)} &-\frac{2P_{(0,0,-1)}}{\sqrt 6}\\
 \end{array}
\right) ,
\end{equation}
where the notation $(I,I_3,S)$ represents the isospin, the third
component of the isospin, and the strange number of the $P_c$ states,
respectively. After expansion, we have
\begin{eqnarray} \label{a3}
&&\mathscr{L}^{8_f}_{eff}=\frac{2}{\sqrt 3}g \big[ \bar P_{(0,1,-1)}\Sigma_z\partial_{z}{\pi^0}P_{(1,0,-1)}+\bar P_{(1,0,-1)}\Sigma_z\partial_{z}{\pi^0}P_{(0,0,-1)} \big]-(f+g)\bar P_{(\frac{1}{2},-\frac{1}{2},0)}\Sigma_z\partial_{z}{\pi^0}P_{(\frac{1}{2},-\frac{1}{2},0)}\nonumber\\
&&+(g+f)\bar P_{(\frac{1}{2},\frac{1}{2},0)}\Sigma_z\partial_{z}{\pi^0}P_{(\frac{1}{2},\frac{1}{2},0)}+(g-f)\bar P_{(\frac{1}{2},-\frac{1}{2},-2)}\Sigma_z\partial_{z}{\pi^0}P_{(\frac{1}{2},-\frac{1}{2},-2)}\nonumber\\
&&+(f-g)P_{(\frac{1}{2},\frac{1}{2},-2)}\Sigma_z\partial_{z}{\pi^0}P_{(\frac{1}{2},\frac{1}{2},-2)}+2f\bar
P_{(1,1,-1)}\Sigma_z\partial_{z}{\pi^0}P_{(1,1,-1)}-2f\bar
P_{(1,-1,-1)}\Sigma_z\partial_{z}{\pi^0}P_{(1,-1,-1)}
\end{eqnarray}
An overall factor $1/f_{\pi}$ is omitted. Using the two independent pionic couplings of the $P_c$ states, we obtain $g$ and $f$. Then, all the other couplings
can be obtained . For the states in the $10_f$ flavor
representation, the Lagrangian is
\begin{eqnarray} \label{a4}
\mathscr{L}^{10_f}_{eff}=g \bar P^{\nu}_c \gamma_{\mu}\gamma_5
\partial^{\mu}\phi P_{c{\nu}}.
\end{eqnarray}
There exists one independent coupling constant.

\end{document}